\documentclass[twocolumn]{aastex631}

\newcommand{\tmop}{\mathrm}
\newcommand{\mathd}{\mathrm{d}}
\shorttitle{TransFit}
\shortauthors{Liu et al.}
\graphicspath{{./}{figures/}}
\usepackage{amsmath}
\usepackage{threeparttable}
\usepackage{booktabs}
\usepackage{changepage} 
\begin{document}
\title{ \texttt{TransFit}: An Efficient Framework for Transient Light-Curve Fitting with Time-Dependent Radiative Diffusion}

\author[0000-0002-8708-0597]{Liang-Duan Liu}
\affiliation{Institute of Astrophysics, Central China Normal University, Wuhan 430079, China; \url{liuld@ccnu.edu.cn;yuyw@ccnu.edu.cn}}
\affiliation{Education Research and Application Center, National Astronomical Data Center, Wuhan 430079, China}
\affiliation{Key Laboratory of Quark and Lepton Physics (Central China Normal University), Ministry of Education, Wuhan 430079, China}

\author{Yu-Hao Zhang}
\affiliation{Institute of Astrophysics, Central China Normal University, Wuhan 430079, China; \url{liuld@ccnu.edu.cn;yuyw@ccnu.edu.cn}}
\affiliation{Education Research and Application Center, National Astronomical Data Center, Wuhan 430079, China}
\affiliation{Key Laboratory of Quark and Lepton Physics (Central China Normal University), Ministry of Education, Wuhan 430079, China}

\author[0000-0002-1067-1911]{Yun-Wei Yu}
\affiliation{Institute of Astrophysics, Central China Normal University, Wuhan 430079, China; \url{liuld@ccnu.edu.cn;yuyw@ccnu.edu.cn}}
\affiliation{Education Research and Application Center, National Astronomical Data Center, Wuhan 430079, China}
\affiliation{Key Laboratory of Quark and Lepton Physics (Central China Normal University), Ministry of Education, Wuhan 430079, China}

\author{Ze-Xin Du}
\affiliation{Institute of Astrophysics, Central China Normal University, Wuhan 430079, China; \url{liuld@ccnu.edu.cn;yuyw@ccnu.edu.cn}}
\affiliation{Education Research and Application Center, National Astronomical Data Center, Wuhan 430079, China}
\affiliation{Key Laboratory of Quark and Lepton Physics (Central China Normal University), Ministry of Education, Wuhan 430079, China}

\author[0009-0004-9719-272X]{Jing-Yao Li}
\affiliation{Institute of Astrophysics, Central China Normal University, Wuhan 430079, China; \url{liuld@ccnu.edu.cn;yuyw@ccnu.edu.cn}}
\affiliation{Education Research and Application Center, National Astronomical Data Center, Wuhan 430079, China}
\affiliation{Key Laboratory of Quark and Lepton Physics (Central China Normal University), Ministry of Education, Wuhan 430079, China}

\author{Guang-Lei Wu}
\affiliation{Institute of Astrophysics, Central China Normal University, Wuhan 430079, China; \url{liuld@ccnu.edu.cn;yuyw@ccnu.edu.cn}}
\affiliation{Education Research and Application Center, National Astronomical Data Center, Wuhan 430079, China}
\affiliation{Key Laboratory of Quark and Lepton Physics (Central China Normal University), Ministry of Education, Wuhan 430079, China}

\author{Zi-Gao Dai}
\affiliation{Deep Space Exploration Laboratory/Department of Astronomy, University of Science and Technology of China, Hefei 230026, People's Republic of China}
\affiliation{School of Astronomy and Space Science, University of Science and Technology of China, Hefei 230026, People's Republic of China}

\begin{abstract}
Modeling the light curves (LCs) of luminous astronomical transients, such as supernovae, is crucial for understanding their progenitor physics, particularly with the exponential growth of survey data. However, existing methods face limitations: efficient semi-analytical models (e.g., Arnett-like) employ significant physical simplifications (like time-invariant temperature profiles and simplified heating distributions), often compromising accuracy, especially for early-time LCs. Conversely, detailed numerical radiative transfer simulations, while accurate, are computationally prohibitive for large datasets. This paper introduces \texttt{TransFit}, a novel framework that numerically solves a generalized energy conservation equation, explicitly incorporating time-dependent radiative diffusion, continuous radioactive or central engine heating, and ejecta expansion dynamics. The model accurately captures the influence of key ejecta properties and diverse heating source characteristics on light curve morphology, including peak luminosity, rise time, and overall shape. Furthermore, \texttt{TransFit} provides self-consistent modeling of the transition from shock-cooling to \(^{56}\mathrm{Ni}\)-powered light curves. By combining physical realism with computational speed, \texttt{TransFit} provides a powerful tool for efficiently inverting LCs and extracting detailed physical insights from the vast datasets of current and future transient surveys.

\end{abstract}


\section{Introduction} \label{sec:intro}

The light curves of luminous transients, including supernovae, kilonovae, fast blue optical transients (FBOTs), and related stellar explosive phenomena, contain critical information about their progenitor origins and underlying physical processes. Given the exponential growth in transient detection rates facilitated by current and next-generation synoptic surveys, particularly the Zwicky Transient Facility (ZTF) \citep{ZTF2019}, the Multi-channel Photometric Survey Telescope (Mephisto) \citep{Yuan2020}, the Wide Field Survey Telescope (WFST) \citep{Wang2023}, Vera C. Rubin Observatory's Legacy Survey of Space and Time (LSST) \citep{LSST2023}, and China Space Station Telescope (CSST) \citep{CSST2019}.  There is an increasing demand for efficient modeling tools to rapidly interpret observed events and distinguish among competing theoretical scenarios.

Semi-analytical models constructed from simplified physical frameworks have proven invaluable for interpreting observational data. In particular, one-dimensional ``Arnett-like” models \citep{Arnett1980,Arnett1982,Kasen2010,Inserra2013} of energy transport in stellar explosions solve the energy conservation equation numerically and derive bolometric light curves through a straightforward numerical integration. This allows efficient fitting of light curves and determination of key explosion parameters, such as ejecta mass, kinetic energy, and energy source characteristics, thereby providing critical insight into the underlying explosion physics \citep{Yu2017,Villar2017,Liu2017,Nicholl2017}. A notable prediction of Arnett-like models is that the luminosity at maximum brightness approximates the instantaneous heating rate of internal energy sources, thereby directly linking observable peak luminosities to critical explosion parameters, particularly the synthesized $^{56}$Ni mass \citep{Stritzinger2006,Drout2011,Scalzo2014}.

Although Arnett-like models are computationally efficient, they rely on significant physical simplifications that limit both their accuracy and their range of applicability. In particular, these models assume from the outset that the spatial distribution of temperature  in the ejecta follows a self-similar solution, implying that the spatial distribution of temperature remains unchanged over time. Numerical simulations, however, indicate that it takes on the order of tens of days for the ejecta to reach such self-similarity in temperature \citep{Khatami2019}. Since the temperature profile governs radiative diffusion and thus the light curves predicted by Arnett-like models often produce inaccuracies in the early-time light curves. Furthermore, the simple integrals used in these models do not capture the impact of the spatial distribution of the heating sources. Moreover, the simplified integral approach employed by Arnett-like models does not account for the spatial distribution of the heating source.  Type Ia supernovae typically exhibit a much higher degree of mixing than core-collapse supernovae, which have a more centrally concentrated $^{56}$Ni \citep{Blondin2013}. Consequently, the simple integral form of Arnett-like models lacking an explicit treatment of the heating source spatial distribution cannot accurately describe the light curves of different supernova classes.

In contrast, detailed radiation‐hydrodynamics simulations—using codes such as \texttt{STELLA} \citep{Blinnikov1993}, \texttt{Sedona} \citep{Kasen2006}, \texttt{CMFGEN} \citep{Hillier2012}, and \texttt{SNEC} \citep{SNEC2015}—explicitly solve the coupled hydrodynamic and radiative‐transfer equations. These models accurately capture time‐dependent diffusion, continuous radioactive heating, and composition‐dependent opacity variations. However, their substantial computational expense renders them impractical for systematic analysis of the large transient samples expected from current and next‐generation surveys.

To bridge the gap between analytic efficiency and numerical completeness, we present a new transient fitting framework called \texttt{TransFit}. Our method is based on a generalized energy conservation equation explicitly incorporating the time-dependent radiative diffusion process, continuous radioactive heating, and ejecta expansion dynamics. The \texttt{TransFit} framework goes significantly beyond traditional semi-analytic approaches by allowing flexibility in modeling diverse spatial density distributions of the ejecta, varying spatial and temporal distributions of radioactive heating sources (e.g., mixed or stratified distributions of  $^{56}$Ni), and compositionally dependent opacity variations. Moreover, our model seamlessly transitions from early-time shock cooling dominated phases to radioactive heating phases, offering a fully self-consistent description of transient luminosity evolution.

By carefully non-dimensionalizing and normalizing the energy conservation equation, we simplified it into a form that is readily amenable to numerical calculations.  Furthermore, we employ an appropriate numerical method to solve the resulting partial differential equations, thereby enabling \texttt{TransFit} achieves computational efficiency comparable to semi-analytic models.  Validation comparisons with sophisticated Monte Carlo radiative transfer simulations (e.g., \texttt{Sedona}) demonstrate excellent agreement across a variety of transient scenarios. By combining physical realism—including sophisticated treatment of the density profiles of the ejecta, radioactive source distributions with computational speed, \texttt{TransFit} offers a powerful and timely tool for extracting detailed insights from the immense datasets produced by next-generation transient surveys.

The paper is structured as follows. Section~\ref{Sec:model} introduces the theoretical foundations of the \texttt{TransFit} framework, including the derivation of key equations and underlying assumptions. Section~\ref{Sec:light_curves} describes the numerical method for computing bolometric light curves and examines the effects of model parameters such as the heating source, ejecta mass, and opacity. In Section~\ref{Sec:Comparison}, we compare \texttt{TransFit} to previous analytical models, highlighting its advantages and limitations. Section~\ref{Sec:Fitting} demonstrates the application of \texttt{TransFit} to fit the light curves of SN\,1993J and SN\,2011kl. Finally, Section~\ref{Sec:Conclusion} summarizes our conclusions and discusses future directions.

\section{Model} \label{Sec:model}

The fundamental principle underlying supernova light-curve modeling is the first law of thermodynamics of an expanding ejecta. The evolution of the light curve is governed by energy deposition, originating either from a shock wave traveling outward or from internal heating mechanisms such as radioactive decay or a central engine. This deposited energy subsequently radiates by diffusing through the optically thick, expanding ejecta, undergoing adiabatic losses before emerging at the photosphere as observable emission.

For a spherically symmetric shell of radius $r$, using the mass $m$ as a Lagrangian coordinate, the fundamental equation can be written as
\begin{equation} \label{Eq:Basic_Eq}
\frac{\partial E}{\partial t} - \frac{P}{\rho^2} \frac{\partial \rho}{\partial t} = \varepsilon_{\mathrm{heat}} - \frac{\partial L}{\partial m},
\end{equation}
where $E$ denotes the specific energy density (i.e., the thermal energy per unit mass). $\rho$ is the density of the ejecta, and $P$ represents the pressure, which in the supernova case, is dominated by radiation, see Appendix~\ref{sec:EOS} for more details. $-\frac{P}{\rho^2}\frac{\partial \rho}{\partial t}$ represents the work done by adiabatic expansion. The radiative cooling term is described by the spherical diffusion equation:
\begin{equation} \label{Eq:dLdm}
\frac{\partial L}{\partial m} =  \frac{1}{4 \pi r^2 \rho} \frac{\partial}{\partial r} (4 \pi r^2 F),
\end{equation}
where $F$ is the radiative flux, which is given by
\begin{equation}
F = - \frac{c}{3 \kappa \rho} \frac{\partial u}{\partial r},
\end{equation}
where $u = \rho E$ is the energy density of the ejecta, $\kappa$ is the opacity of the ejecta, and $c$ is the speed of light.

The local energy generation rate of radiation, $\varepsilon_{\tmop{heat}}$,
denotes the heating rate per gram per second. It can be generally expressed as
\begin{equation}
    \varepsilon_{\rm{heat}} (r, t) = \varepsilon_{\rm{heat} 0}
\xi_{\rm{heat}} (r) f_{\rm{heat}} (t),
\end{equation}
where $\varepsilon_{\tmop{heat} 0}$ is the amplitude of the specific power
input (in erg g$^{- 1}$ s$^{- 1}$), $\xi_{\tmop{heat}} (r)$ is its radial
distribution, and $f_{\rm{heat}} (t)$ describes its time dependence.

We assume that the ejecta expand homologously, and thus their outer radius evolves according to
\begin{equation}
    R_{\rm max} = R_0 + v_{\rm max} t,
\end{equation}
where $v_{\rm max}$ is the maximum ejecta velocity and $R_0$ is the radius of the SN progenitor. We introduce a comoving, dimensionless radial coordinate $x$ to follow the expansion of matter, defined as
\begin{equation}
    x \equiv \frac{r}{{R_{\rm{max}}} },
\end{equation}
where $r$ is the distance of a specific layer from the center of the
explosion. For all elements of mass within the ejecta, $0 \leq x \leq 1$. Under homologous expansion, the ejecta density is given by
\begin{equation}
    \rho  (r, t) = \rho_0 \left( \frac{R_0}{R_{\rm max}} \right)^3 \eta_{\rm{{ej}}} (x),
\end{equation}
where $\rho_0$ is the characteristic density of the ejecta, the factor $\left( {R_0}/{R_{\rm max}} \right)^3$  accounts for the expansion, and
$\eta_{\tmop{ej}} (x)$ describes the density profile, normalized such that
$\eta_{\tmop{ej}} (0) = 1$.

Assuming that radiative cooling and source heating are negligible, the internal energy evolves solely due to adiabatic expansion. In a purely adiabatic process, the internal energy density scales as $u \propto
R_{\max}^{- 4}$. To remove this dependency, we define the internal energy density as
\begin{equation}
u(r, t) = u_0 \left( \frac{R_0}{R_{\rm{max}}} \right)^4 e (r, t),
\end{equation}
where $u_0$ is the characteristic internal energy density, $\left( {R_0}/{R_{\rm max}} \right)^4$ accounts for the decrease in internal energy density due to adiabatic expansion, and $e(r,t)$ represents the dimensionless internal energy density distribution over time and radius.

In order to simplify the fundamental equations, we define a characteristic diffusion timescale as
\begin{equation}
t_{\tmop{diff}} \equiv \frac{3 \kappa \rho_0 R_0^2}{c},
\end{equation}
and we nondimensionalize the time variable using this timescale, i.e.,
\begin{equation}
    y \equiv \frac{t}{t_{\tmop{diff}}}.
\end{equation}
After dimensionless processing both the temporal and spatial variables, the fundamental equation Eq (\ref{Eq:Basic_Eq})  reduces to
\begin{equation}
\frac{\partial e (x, y)}{\partial y} =  \frac{1}{x^2}
\frac{\partial}{\partial x} \left[ D (x, y) \frac{\partial e (x,
y)}{\partial x} \right] + S (x, y).
\end{equation}
Here, the left-hand side represents the time derivative of the energy density, the first term on the right-hand side corresponds to the radial diffusion, and the second term represents a source term accounting for local heating. The dimensionless, position- and time-dependent diffusion coefficient,
$ D$,  is defined as
\begin{equation}
    D (x, y) \equiv \frac{x^2}{\eta_{\tmop{ej}} (x)} \left(
\frac{R_{\max}}{R_0} \right),
\end{equation}
and the dimensionless heat generation in the system is described by
\begin{equation}
    S (x, y) \equiv \eta_{_{\tmop{ej}}} (x) \xi_{\tmop{heat}} (x)
f_{\rm{heat}} (y) \left( \frac{R_{\max}}{R_0} \right).
\end{equation}
To directly solve this partial differential equation using numerical methods, it is necessary to specify the boundary and initial conditions.

\subsection{Initial Conditions}

The initial conditions for supernova light-curve modeling depend on the evolutionary history and final explosion of massive stars \citep{Woosley2002,Sukhbold2016}. Typically, these models begin by specifying the ejecta mass, composition (including radioactive elements), and the density and velocity profiles immediately after the explosion.

In our calculations, we define the initial time as the moment when the shock front reaches the progenitor's surface at $r=R_0$. By this point, the ejecta has been heated and accelerated. The strong shock approximately divides the total energy of the supernova ejecta into thermal and kinetic components  ($E_{\rm{Th}}$ and $E_{\rm K}$) forms,  so the total supernova energy can be written as
\begin{equation}
E_{\mathrm{SN}}=E_{\mathrm{Th}}+E_{\mathrm{K}}.
\end{equation}

The initial thermal energy is
\begin{equation}
E_{\mathrm{Th,in}}  =\int_0^{R_0} 4 \pi r^2 a u \mathrm{~d} r =4 \pi R_0^3 u_0 I_{\mathrm{Th}},
\end{equation}
where
\begin{equation}
    I_{\tmop{Th}} \equiv \int_0^1 x^2 e (x, 0) \text{d} x,
\end{equation}
is a dimensionless thermal energy factor, and $e(x,0)$ denotes the dimensionless  initial energy density distribution.

During adiabatic, homologous expansion, the thermal energy evolves as
\begin{equation}
    E_{\rm Th}(t) = E_{\rm Th,in} \left( \frac{R_0}{R_{\rm max}} \right),
\end{equation}
gradually converting into kinetic energy. This conversion leads to an approximate 40$\%$ increase in velocity \citep{Arnett1980}. Most of this conversion occurs within the initial expansion timescale \( t_{\rm ex} = R_0 / v_{\rm max} \), which justifies the assumption of a nearly constant velocity at later times. For a given density profile, the total kinetic energy is
\begin{equation}
E_{\text{K}} = \int_0^{R_{\max} } \frac{1}{2} \rho v^2 4 \pi r^2 \text{d} r = 2\pi \rho_0 R_0^3 v_{\tmop{max}}^2 I_{\text{K}},
\end{equation}
where 
\begin{equation}
    I_{\rm K} \equiv \int_0^1 x^4 \eta_{\rm ej}(x) \text{d} x,
\end{equation}
is a dimensionless factor for kinetic energy.

The total mass of the ejecta follows from integrating the density profile:
\begin{equation}
    M_{\tmop{ej}} = \int_0^{R_0} 4 \pi r^2 \rho \text{d} r = 4 \pi R_0^3 \rho_{0} I_{\rm M},
\end{equation}
where 
\begin{equation}
    I_{\rm M} \equiv \int_0^1 x^2 \eta_{\rm ej}(x) \text{d} x,
\end{equation}
is a dimensionless factor for the ejecta, which is determined by the assumed density profile. For a uniform density, $I_{\rm M}=1/3$.

Initially, the ejecta is optically thick, with an optical depth
\begin{equation}
    \tau_{ 0}= \int_0^{R_0} \kappa \rho_0 \eta_{\tmop{ej}} (x) \text{d} r = \kappa_0 \rho_0 R_0 I_{\tau},
\end{equation}
where 
\begin{equation}
    I_{\tau} \equiv \int_0^1  \eta_{\rm ej}(x) \text{d} x,
\end{equation}
is a dimensionless factor for the optical depth of ejecta.
Due to the homologous expansion, the optical depth decreases over time according to
\begin{equation}
\tau(t) = \tau_0 \left( \frac{R_0}{{R_{\tmop{max}}} } \right)^2 .
\end{equation}
When $\tau$ reaches approximately unity, the entire ejecta
becomes transparent, permitting photons produced anywhere within it to escape
with minimal scattering. This transition marks the onset of the nebular phase.
\begin{figure*}[ht!]
  \centering
  \includegraphics[width=0.90\textwidth]{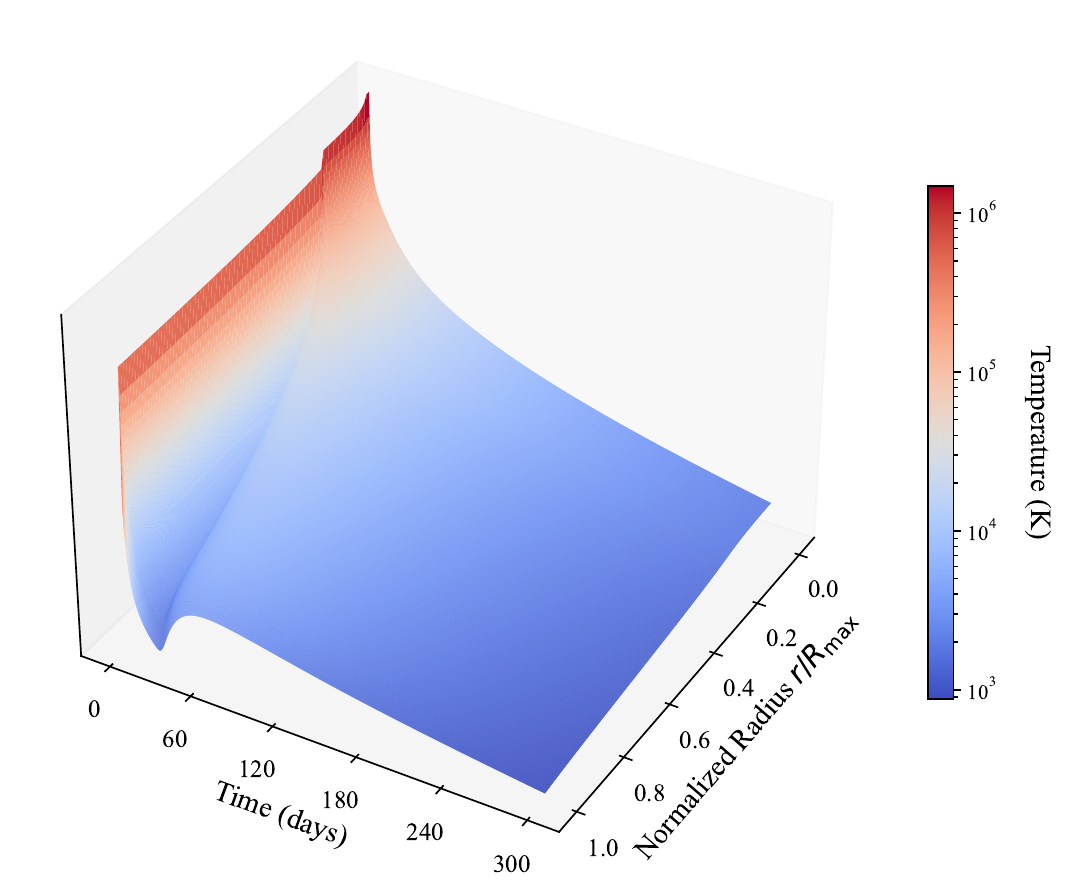}
  \caption{Temporal and spatial evolution of the ejecta temperature \(T\)  as a function of normalized radius \(r/R_{\max}\) and time.  }
  \label{Fig:Temperature-3d}
\end{figure*}

\subsection{Boundary Conditions}

For most Type Ia and core-collapse supernovae after shock breakout, the
primary energy source fueling the light curve is the radioactive decay of
$^{56} \tmop{Ni}$ to $^{56} \tmop{Co}$ and subsequently to $^{56} \tmop{Fe}$.
The radioactive material may be distributed throughout the ejecta or confined to a specific region.  In a spherically symmetric scenario, no additional photon source is assumed at the bottom of the ejecta $(r = R_{\min})$. Consequently, the
flux there is set to zero $F=0$, which is equivalent to
\begin{equation}
\left.\frac{\partial e(x,y)}{\partial x}\right|_{x=x_{\min}}=0.
\end{equation}

For supernovae with a centrally located energy source, such as magnetar
spin-down or fallback accretion, the heating arises from the innermost region
of the ejecta. Consequently, photons must diffuse through the entire ejecta
before becoming observable. In this scenario, the radiative flux at the inner
boundary $(r = R_{\min})$ depends on the central heating rate. The inner
boundary condition for the radiative flux  is specified as
\begin{equation}
F(R_{\min},t)=\frac{L_{\mathrm{engine}}(t)}{4\pi R_{\mathrm{min}}^{2}},
\end{equation}
where $L_{\tmop{engine}} (t)$ is the time-dependent luminosity generated by
the central engine (magnetar spin-down power or fallback accretion). As a result, the source term in the diffusion equation is incorporated into this inner boundary condition as
\begin{equation}
   \left.\frac{\partial e(x,y)}{\partial x}\right|_{x=x_{\min}} = f_{\tmop{ib}} (y),
\end{equation}
where $ f_{\tmop{ib}} (y)$ is a time-dependent function specifying the inner boundary condition for a centrally located heating source.

The outer boundary condition governs the transition of photons from the optically thick interior to free escape at the photosphere or the outer edge of the ejecta. At the surface, we adopt the plane-parallel, gray-atmosphere approximation, assuming the surface layer's thickness is negligible relative to the ejecta radius. This approach, referred to as the Eddington surface boundary condition, is expressed as
\begin{equation}
    T^4 = \frac{3}{4} T_{\tmop{eff}}^4 \left( \tau + \frac{2}{3} \right),
\end{equation}
where  $\frac{2}{3}$ comes from the Eddington approximation, $T_{\tmop{eff}}$ denotes the effective temperature and $\tau$ is the
optical depth.  At the outer edge of the ejecta, $\tau=0$, and we have $T^{4} (x_{\max},y)=\frac{1}{2}T_{\mathrm{eff}}^{4}(y).$

In supernova ejecta, where radiation dominates, the energy density can be written as $u = a T^4$ , with $a$ representing the radiation constant. The boundary condition is then
\begin{equation}
    e (x_{\max}, y) =f_{\tmop{ob}} (y)  \left.\frac{\partial e(x,y)}{\partial x}\right|_{x=x_{\max}},
\end{equation}
where $ f_{\tmop{ob}} (y)$ is a time-dependent function specifying the outer boundary condition,
\begin{equation}
    f_{\tmop{ob}} (y) \equiv - \frac{2}{3 \kappa \rho_0 R_0} 
\frac{1}{\eta_{\tmop{ej}} (x_{\max})} \left( \frac{R_{\max}}{R_0} \right)^2.
\end{equation}

According to the Eddington approximation, the photospheric radius
$R_{\tmop{ph}}$ is defined by the condition $\tau = 2 / 3$:
\begin{equation}
\int_{R_{\mathrm{ph}}}^{R_{\rm max}} \kappa \rho \text{d} r=\frac{2}{3}.
\end{equation}
In dimensionless form,  this condition is expressed as
\begin{equation}
  \int_{x_{\tmop{ph}}}^1 \eta_{\rm ej} (x) \text{d} x = \frac{2}{3} \left(
\frac{R_{\tmop{max}}}{R_0} \right)^2 \frac{1}{\kappa  \rho_0 R_0},  
\end{equation}
where $x_{\tmop{ph}} \equiv R_{\tmop{ph}} / R_{\max}$. Given a specific ejecta
density profile, one can numerically determine the time evolution of
$x_{\tmop{ph}}$. As expansion continues, $x_{\tmop{ph}}$ decreases, reflecting
a recession of the photosphere in comoving coordinates. Consequently, the
evolution of $R_{\tmop{ph}}$ is governed by the competing effects of ejecta
expansion and photospheric recession. Indeed, \cite{Liu2018}  demonstrated that, independent of the assumed density distribution, $R_{\tmop{ph}}$ initially increases and subsequently declines at later times,
this result that is robust to the details of radiative transfer and cooling processes within the ejecta.

The luminosity at the outer edge of the ejecta is
\begin{equation}
L_{\tmop{bol}} (x_{\max}, y) = - L_0 \left[ \frac{x^2}{\eta_{\tmop{ej}} (x)}
\frac{\partial e (x, y)}{\partial x} \right]_{x = x_{\max}},
\end{equation}
where the negative sign indicates that the luminosity is associated with the outflow of radiation from regions of higher to lower energy density, $L_0 \equiv 4 \pi R_0^3 u_0 / t_{\tmop{diff}}$ is the characteristic bolometric luminosity.

\section{Numerical Light Curves} \label{Sec:light_curves}

A complete description of the numerical procedure is provided in Appendix~\ref{sec:Numerical Implementation}. The theoretical light curves critically depends on two sets of model input parameters:

\begin{itemize}
    \item \textbf{Ejecta parameters:}
          ejecta mass \(M_{\mathrm{ej}}\); 
          density profile \(\eta_{\mathrm{ej}}(x)\);
          initial radius \(R_0\); 
          initial kinetic energy \(E_{\mathrm{K}}\); 
          initial thermal energy \(E_{\mathrm{Th,in}}\); 
          gray opacity \(\kappa\); 
          and initial internal-energy distribution \(e(x,0)\).
    \item \textbf{Heating/source-term parameters:}
          specific-power amplitude \(\varepsilon_{\mathrm{heat}0}\);
          radial heating profile \(\xi_{\mathrm{heat}}(r)\); 
          and temporal heating function \(f_{\mathrm{heat}}(t)\).
\end{itemize}

These parameters jointly determine the hydrodynamic structure of the ejecta and the rate of energy deposition, thereby controlling the shape and luminosity scale of the resulting light curve. In our calculations, we adopt the following default values: an ejecta mass of \(M_{\mathrm{ej}} = 5.0\,M_\odot\) with a uniform density profile \(\eta_{\mathrm{ej}}(x)=1\); initial radius \(R_0=100R_{\odot}\); a kinetic energy of \(E_{\mathrm{K}} = 10^{51}\,\mathrm{erg}\); an initial thermal energy of \(E_{\mathrm{Th,in}} = 10^{49}\,\mathrm{erg}\); an opacity of \(\kappa = 0.2\,\mathrm{cm}^2\,\mathrm{g}^{-1}\); a synthesized nickel mass of \(M_{\mathrm{Ni}} = 0.2\,M_\odot\); and a maximum dimensionless heating radius of \(x_{\mathrm{heat}} = 0.2\).

Temperature gradients within the ejecta drive its spectral evolution and thus the color changes observed in the light curve. Accurate modeling of this temperature evolution is essential for interpreting both photometric and spectroscopic features of the supernova. Figure~\ref{Fig:Temperature-3d} shows the ejecta temperature as a function of normalized radius and time on a logarithmic color scale (from \(10^6\)\,K in red to \(10^3\)\,K in blue). Initially, the ejecta is nearly isothermal and extremely hot; thereafter it cools from the outer layers inward via radiative diffusion and adiabatic expansion. By \(\sim 100\)\,days, temperatures have dropped to \(10^4\)–\(10^5\)\,K, and by \(\sim 300\)\,days to \(10^3\)–\(10^4\)\,K. 

We explore a range of ejecta and heating-source properties to capture the diversity of observed supernova light curves.

\subsection{Dependence on Heat Source}

The radioactive decay of elements synthesized during explosive nucleosynthesis—particularly \(^{56}\mathrm{Ni}\)—is the primary power source for the light curves of many ordinary supernovae. In typical core‐collapse supernovae, the \(^{56}\mathrm{Ni}\) mass lies in the range \(0.01\)–\(0.1\,M_\odot\), whereas Type Ia and energetic, gamma ray burst associated Type Ic supernovae can synthesize up to \(0.6\,M_\odot\) of \(^{56}\mathrm{Ni}\) \citep{Cano2017,Kasen2017}.

The mass of \(^{56}\mathrm{Ni}\) synthesized in the supernova shock is constrained by the explosion energy and the pre-shock stellar density. Production of \(^{56}\mathrm{Ni}\) requires temperatures exceeding \(T_{\mathrm{Ni}}\approx5\times10^9\)\,K. Assuming spherical symmetry, the maximum radius within which the post-shock temperature can reach \(T_{\mathrm{Ni}}\) is \citep{Woosley1986}:
\begin{equation}
   R_{\mathrm{Ni}}\approx\left(\frac{3E_{\mathrm{SN}}}{4\pi a\,T_{\mathrm{Ni}}^4}\right)^{1/3}
\simeq 3700\;\mathrm{km}\,\left(\frac{E_{\mathrm{SN}}}{10^{51}\,\mathrm{erg}}\right)^{1/3}. 
\end{equation}

\begin{figure}
    \centering
\includegraphics[width=0.45\textwidth]{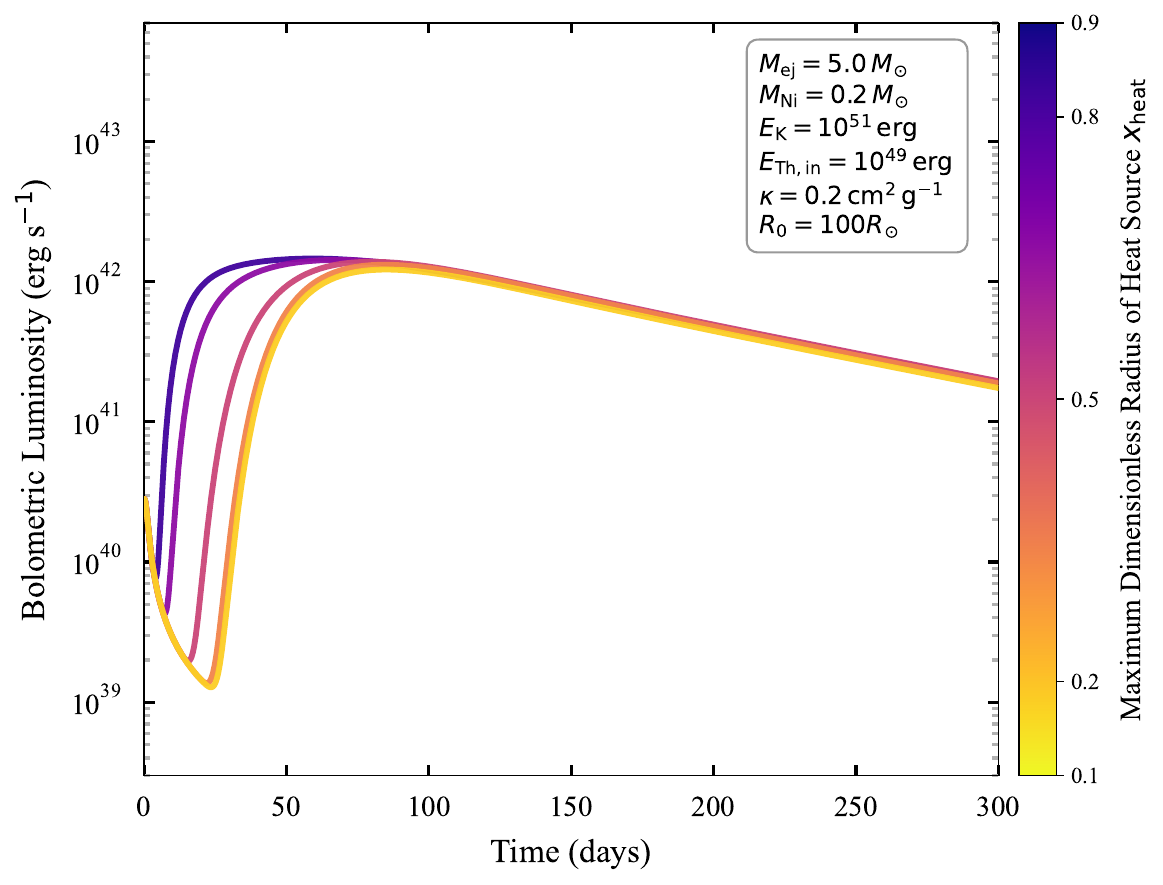}
    \caption{ Impact of the heat source distribution, the heating is uniformly mixed out to \(x_{\mathrm{heat}}\),  on supernova bolometric lightcurves. 
}
    \label{Fig:lc_different_x_s}
\end{figure}

The degree of \(^{56}\mathrm{Ni}\) mixing critically influences the early lightcurve  by modulating the efficiency with which decay gamma-rays heat the ejecta. Weak mixing---characterized by an inner mixing boundary deep within the ejecta---concentrates nickel near the center, whereas strong mixing produces a more uniform nickel distribution. SN~Ia typically exhibit a much larger degree of mixing, while core-collapse SNe tend to have more centrally concentrated \(^{56}\mathrm{Ni}\) \citep{Blondin2013}. For simplicity, we model the spatial distribution of this nickel heating source within the ejecta as a step function:
\begin{equation} \label{eq:heating_profile}
 \xi_{\mathrm{heat}}(x) =
 \begin{cases}
  \xi_0, & x_{\min} \le x \le x_{\mathrm{heat}},\\
  0,    & x_{\mathrm{heat}} < x \le x_{\max},
 \end{cases} 
\end{equation}
where \(\xi_0\) is a normalization constant ensuring the correct total amount of \(^{56}\mathrm{Ni}\), \(x_{\min}\) and \(x_{\max}\) denote the inner and outer dimensionless ejecta boundaries, respectively, and \(x_{\mathrm{heat}}\) represents the dimensionless mixing radius, controlling the extent of the heated region. Using this model, we explore how the spatial distribution of the heating source, parameterized by \(x_{\mathrm{heat}}\), impacts the resulting light curves. As shown in Figure~\ref{Fig:lc_different_x_s}, models with weak mixing (small \(x_{\mathrm{heat}}\)) trap energy deep in the ejecta, producing broader, fainter peaks that occur later, whereas strong mixing (large \(x_{\mathrm{heat}}\)) shifts heating to shallower layers, yielding brighter maxima reached more quickly and followed by a steeper early decline. Despite these differences, all curves converge at late times onto the same \(^{56}\mathrm{Co}\) decay tail, indicating that competition between diffusion and heating processes, rather than the total nickel mass, governs the shape of the lightcurve peak.

\begin{figure}
    \centering
\includegraphics[width=0.45\textwidth]{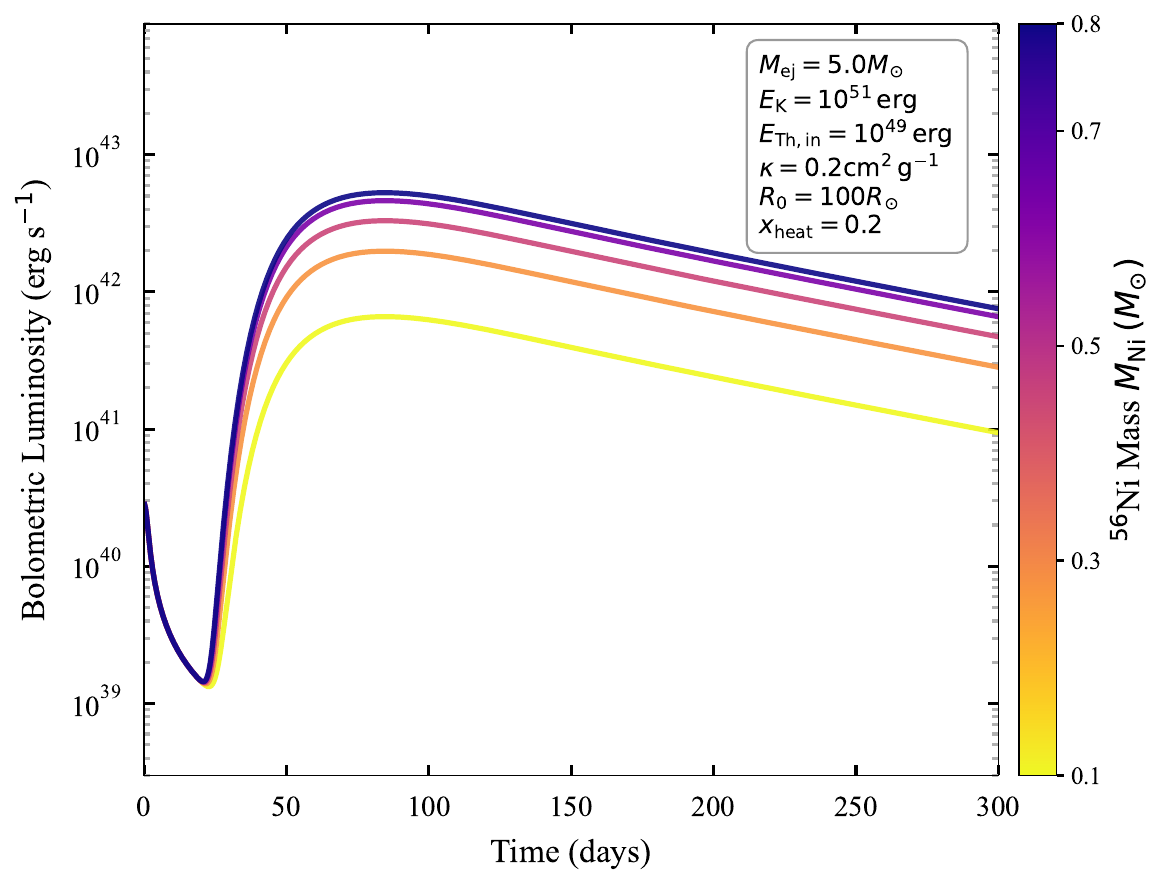}
    \caption{ Impact of the mass of $^{56}$Ni
    $M_{\mathrm{Ni}}$,  on supernova bolometric lightcurves.
}
    \label{Fig:lc_different_M_Ni}
\end{figure}

The amount of radioactive material produced in a SN explosion is a key factor that influences the transient behavior of all SN types.  Figure \ref{Fig:lc_different_M_Ni} illustrates how the amount of $^{56}$Ni synthesized in a supernova explosion affects the resulting bolometric lightcurve. The early phase of the bolometric light curve is primarily powered by the diffusion of the initial internal energy stored within the ejecta. Consequently, the synthesized mass of \(^{56}\mathrm{Ni}\) is not expected to significantly influence the light curve during this initial period. However, a higher mass of synthesized \(^{56}\mathrm{Ni}\) leads to a pronounced increase in both the peak luminosity and the brightness of the subsequent radioactive decay tail.

Explaining the extreme luminosities of superluminous supernovae (SLSNe) purely via radioactive decay would demand implausibly large \(^{56}\mathrm{Ni}\) masses, on the order of several solar masses \citep{Gal-Yam2019}. An alternative mechanism invokes a long-lived central engine that continuously injects energy into the ejecta over months to years \citep{Kasen2010,Woosley2010}. Soon after pulsars were identified as rapidly rotating, magnetized neutron stars, \cite{Ostriker1969} proposed that magnetic-dipole spin-down could power supernova light curves \footnote{ Stable millisecond magnetars formed in the aftermath of binary‑neutron‑star mergers can also power bright optical and X‑ray emission\citep{Yu2013,Metzger2014,Yu2018}.}. A massive star may collapse into a neutron star with an initial spin period \(P_{\mathrm{i}}\) of order milliseconds. The rotational kinetic energy of such a neutron star is
\begin{equation} \label{eq:Erot}
E_{\mathrm{rot}} = \frac{I_{\mathrm{NS}}}{2} \left( \frac{2 \pi}{P_{\mathrm{i}}} \right)^2 \simeq 2.5 \times 10^{52} P_{\mathrm{ms}}^{-2} \left( \frac{M_{\mathrm{NS}}}{1.4 M_\odot} \right)^{3/2} \,\mathrm{erg},
\end{equation}
where \(P_{\mathrm{ms}} = P_{\mathrm{i}} / (1\,\mathrm{ms})\) defines the initial period in milliseconds, \(M_{\mathrm{NS}}\) is the neutron star mass, and the moment of inertia of the neutron star is taken as \(I_{\mathrm{NS}} \simeq 1.3 \times 10^{45} (M_{\mathrm{NS}} / 1.4 M_\odot)^{3/2}\,\mathrm{g}\,\mathrm{cm}^2\).

\begin{figure*}
    \centering
\includegraphics[width=0.45\textwidth]{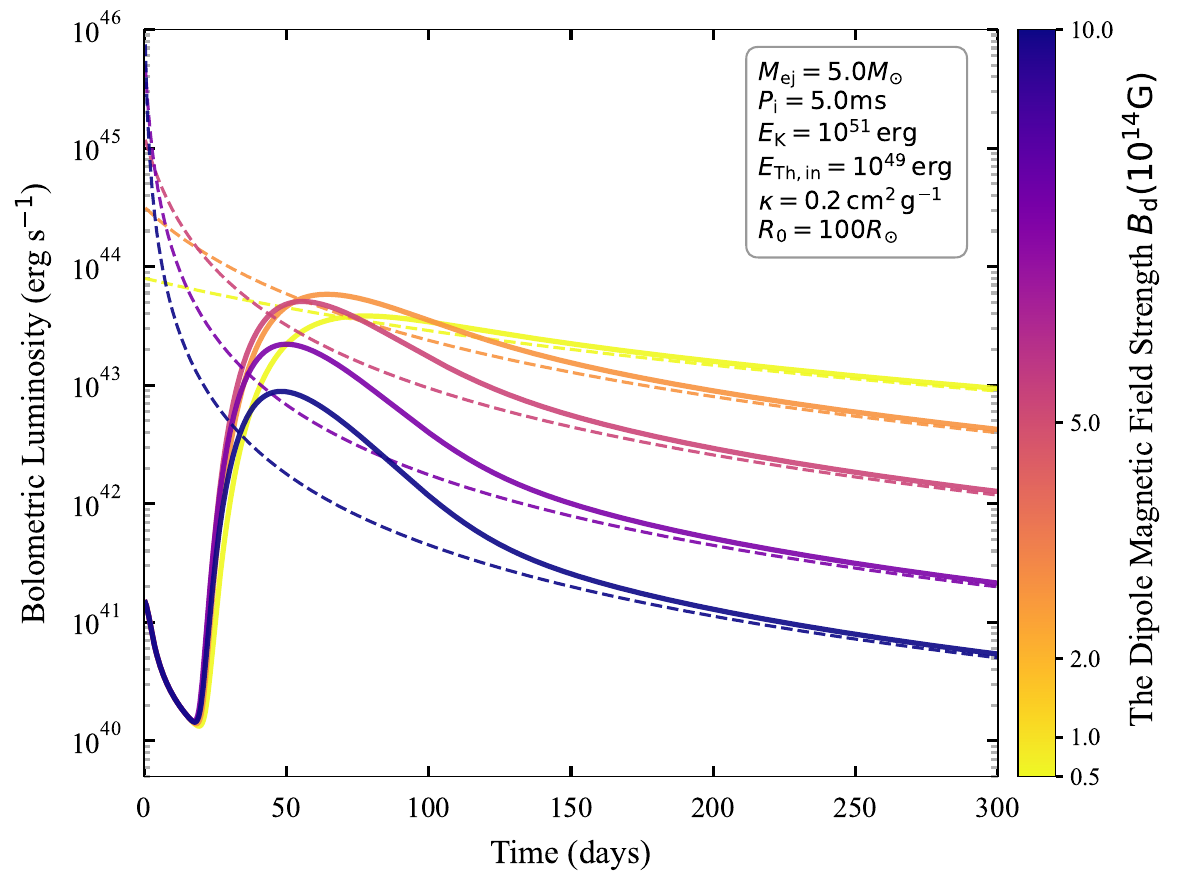}
\includegraphics[width=0.45\textwidth]{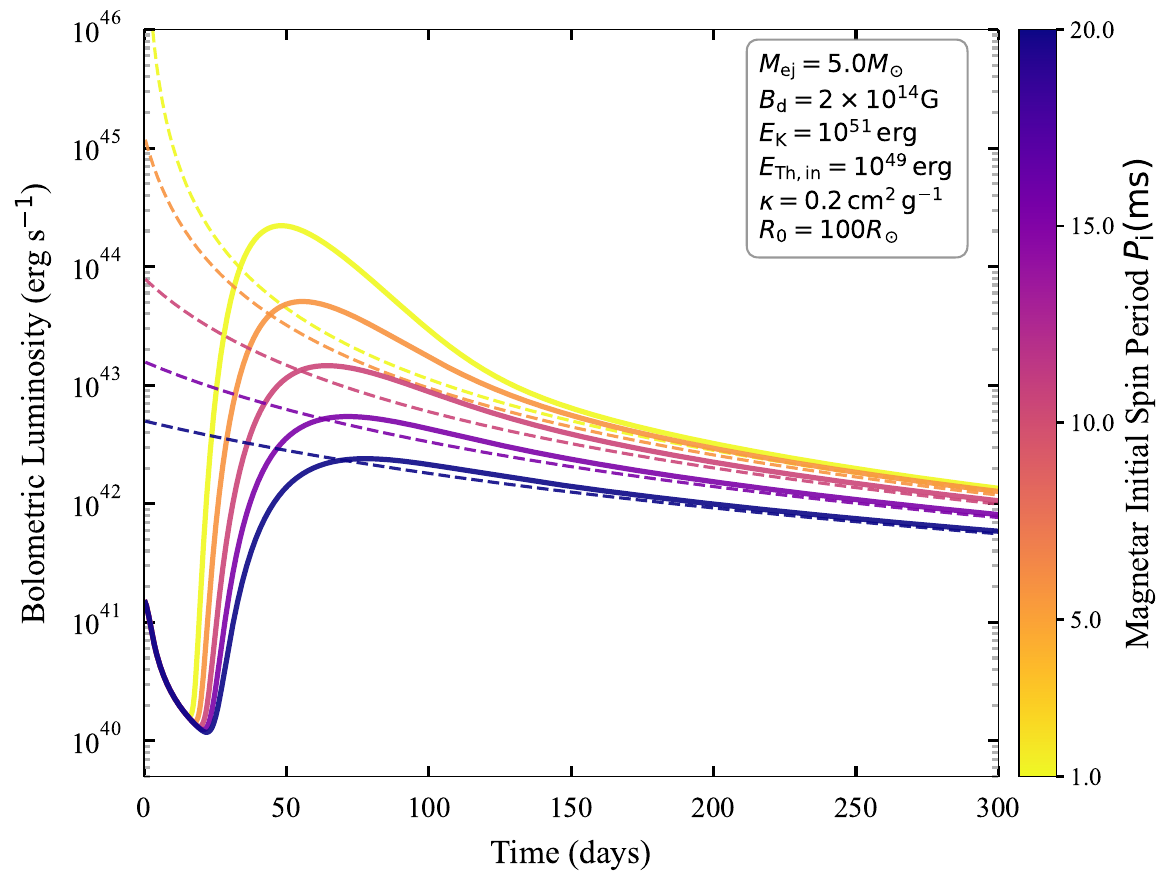}
    \caption{ Impact of the 
   the dipole magnetic field strength $B_\mathrm{d}$ and magnetar initial spin period $P_{\rm{i}}$,  on supernova bolometric lightcurves. The dashed lines represent the central‐engine heating rate.
}
    \label{Fig:lc_different_P_B}
\end{figure*}

The spin-down luminosity provided by magnetic dipole radiation evolves as:
\begin{equation} \label{eq:Lsd}
   L_{\mathrm{sd}}(t) = L_{\mathrm{sd,i}}\biggl(1 + \frac{t}{t_{\mathrm{sd}}}\biggr)^{-2}.
\end{equation}
Here, \(L_{\mathrm{sd,i}} = E_{\mathrm{rot}} / t_{\mathrm{sd}}\) is the initial spin-down luminosity, and \(t_{\mathrm{sd}}\) is the characteristic spin-down timescale, given by
\begin{equation} \label{eq:tsd}
 t_{\mathrm{sd}} \simeq 0.5 \left( \frac{B_{\mathrm{d}}}{10^{14}\,\mathrm{G}} \right)^{-2} \left( \frac{P_{\mathrm{i}}}{1\,\mathrm{ms}} \right)^2 \left( \frac{M_{\mathrm{NS}}}{1.4 M_\odot} \right)^{3/2} \left( \frac{R_{\mathrm{NS}}}{10\,\mathrm{km}} \right)^{-6} \,\mathrm{days},
\end{equation}
where \(B_{\mathrm{d}}\) denotes the surface dipole magnetic field strength at the neutron star pole, and \(R_{\mathrm{NS}}\) is the neutron star radius.

In central engine model, the heating source is typically assumed to remain at the center of the ejecta, requiring the injected energy to diffuse outwards. Figure~\ref{Fig:lc_different_P_B} illustrates how variations in the dipole magnetic field strength $B_\mathrm{d}$ and the magnetar’s initial spin period $P_{\rm{i}}$ affect the bolometric light curves. A stronger dipole field strength $B_\mathrm{d}$  increases the rate of rotational‐energy injection: the initial spin‐down luminosity scales as $L_{\text{sd,i}} \propto B_{\text{d}}^2$, while the spin‐down timescale scales as $t_{\tmop{sd}} \propto B_{\text{d}}^{- 2}$. Consequently, energy is deposited earlier, producing a faster rise to peak luminosity and a more rapid post‐peak decline as the central engine exhausts its reservoir more quickly. A shorter initial spin period supplies more rotational energy and a higher early injection rate, yielding a brighter peak luminosity. Because energy is deposited more rapidly at early times, models with smaller $P_{\rm{i}}$ exhibit a faster rise and attain their peak luminosity sooner.

To power the observed supernova light curve, the energy released by magnetar spin-down must be efficiently thermalized within the ejecta. However, the efficiency of this thermalization process is not well constrained.  The spin-down energy released from a magnetar is thought to be initially in the form of a Poynting flux, subsequently transported by an ultra-relativistic electron-positron wind \citep{Gaensler2006}. When this magnetar wind catches up with and collides with the slower-moving supernova ejecta, a reverse shock can form within the wind, converting its bulk kinetic energy into thermal energy. The hot electron-positron plasma in the reverse-shocked wind  then releases its internal energy primarily through synchrotron emission and inverse-Compton scattering by these relativistic leptons. Consequently, the photons emitted from the magnetar wind nebular are predominantly X-rays \citep{Metzger2014, Yu2019}, necessitating the inclusion of X-ray opacity when calculating the relevant absorption coefficients.  Producing the high luminosities characteristic of SLSNe via this mechanism requires both rapid initial rotation (millisecond periods) and extremely high magnetic fields (\(B_\mathrm{d} \gtrsim 10^{14}\,\mathrm{G}\)) for the neutron star \citep{Yu2017,Liu2017,Nicholl2017}.

\subsection{Dependence on Ejecta Mass and Kinetic Energy}

\begin{figure*}
    \centering
\includegraphics[width=0.45\textwidth]{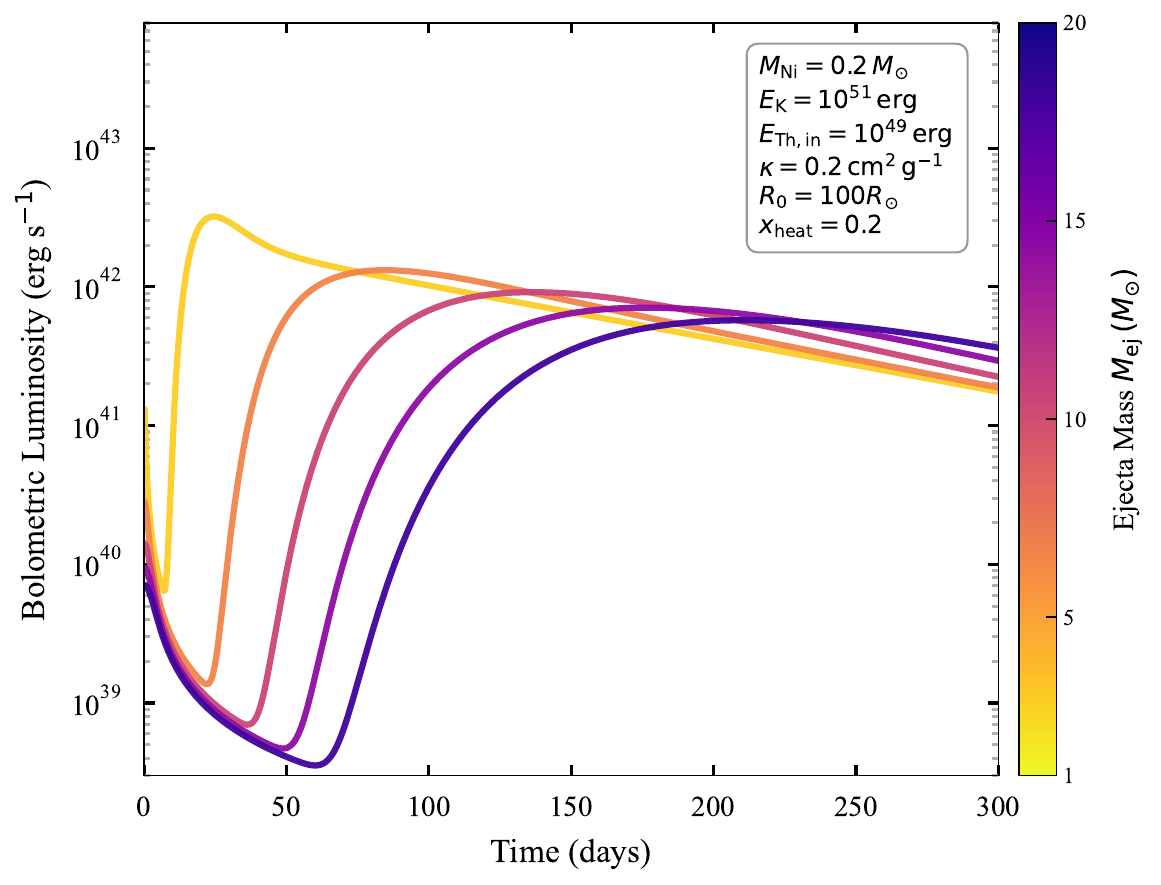}
\includegraphics[width=0.45\textwidth]{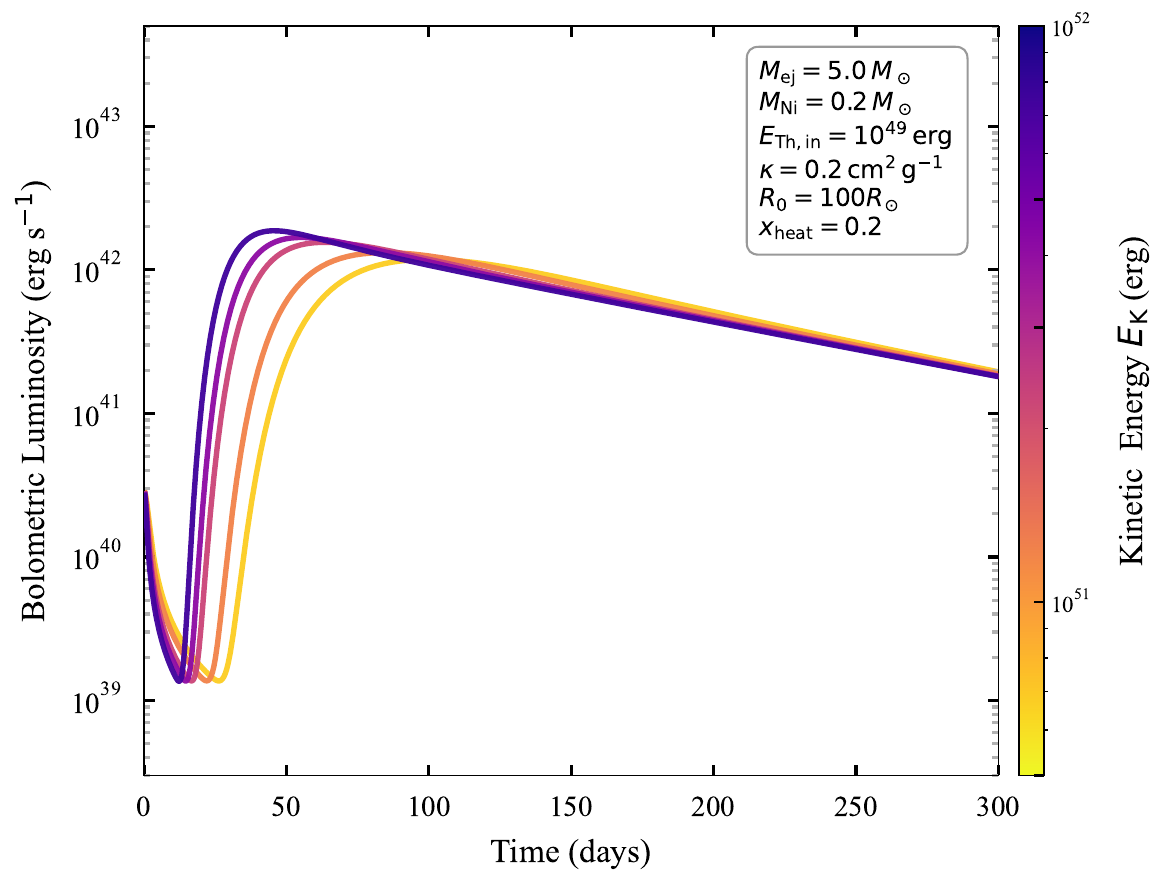}
    \caption{Impact of the 
   the ejecta mass $M_\mathrm{ej}$ and initial kinetic energy $E_{\rm{K}}$,  on supernova bolometric lightcurves.}
    \label{Fig:lc_different_M_ej_and_E_K}
\end{figure*}

Figure~\ref{Fig:lc_different_M_ej_and_E_K} illustrates how the ejecta mass
\(M_{\mathrm{ej}}\) and the initial kinetic energy $E_{\rm{K}}$ shape the bolometric
light curve. For a fixed \(^{56}\mathrm{Ni}\) mass, a larger
\(M_{\mathrm{ej}}\) spreads the same energy over more material, lengthening the
photon diffusion timescale and increasing adiabatic losses before maximum light,
which lowers the peak luminosity. In core‑collapse events, however, more massive
progenitors often synthesize additional \(^{56}\mathrm{Ni}\), partially
offsetting this effect. In magnetar‑powered models, a larger
\(M_{\mathrm{ej}}\) raises the thermalization mass and photon‑trapping time, and
can boost the peak luminosity when the engine’s injection timescale matches the
extended diffusion timescale. By contrast, a higher $E_{\rm{K}}$ shortens the diffusion timescale, and makes the ejecta optically thin
earlier, leading to a faster rise, a narrower peak.

\subsection{Dependence on Ejecta Opacity}

A key simplification in semi-analytic light-curve models is the use of a constant Thomson-scattering opacity, taken as the average ejecta opacity. In more realistic situations that are accounted for in radiation hydrodynamics models, opacity is space and time dependent. This treatment assumes perfectly elastic photon–electron interactions, thereby neglecting net momentum transfer. In a fully ionized plasma, the resulting gray opacity is frequency-independent and well approximated by
\begin{equation}
  \kappa_{\rm es}\simeq0.2\,(1+X)\;\mathrm{cm}^2\,\mathrm{g}^{-1},  
\end{equation}
where \(X\) is the hydrogen mass fraction. For a solar-composition mixture (\(X\simeq0.70\)), this yields \(\kappa_{\rm es}\approx0.34\)\,\(\mathrm{cm}^2\,\mathrm{g}^{-1}\). This opacity is critical for modeling radiation transport in ionized astrophysical environments, such as the interiors of hot stars, where electron scattering dominates. In rapidly expanding ejecta, line opacity can become comparable to or exceed
electron scattering opacity , which complicates analytical light curve
modeling. because of the rapidly
changing opacities in their ejecta, the calculated average
opacities show reasonably good agreement with frequently
used constant opacities  \citep{Nagy2018} A common, albeit simplistic, approach to account for line opacity is
to impose a minimum value on the total opacity. In previous works, minimum
opacity values ranging between 0.01 and 0.34 $\mathrm{cm}^2 \hspace{0.17em}
\mathrm{g}^{- 1}$ have been used.

\begin{figure}
    \centering
\includegraphics[width=0.45\textwidth]{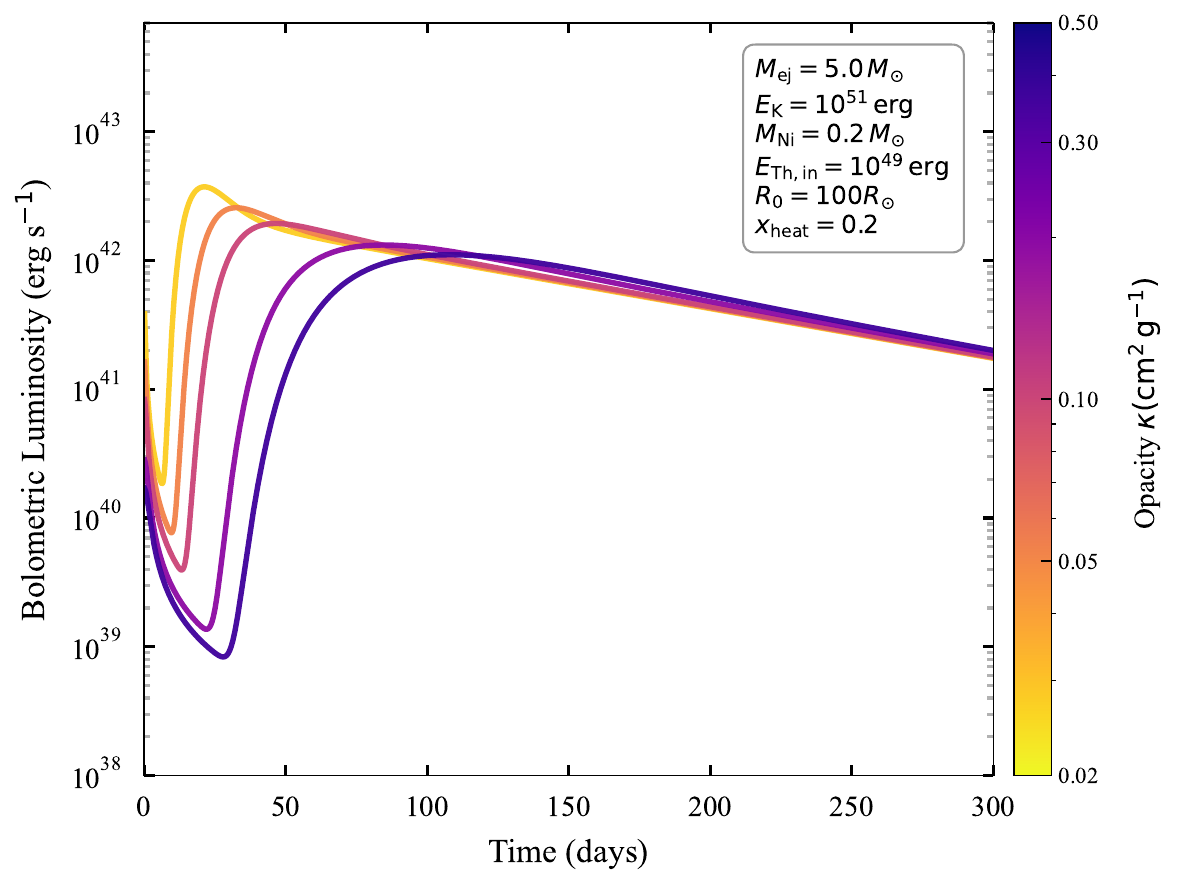}
    \caption{ Impact of the opacity
    $\kappa$,  on supernova bolometric lightcurves.
}
    \label{Fig:lc_different_kappa}
\end{figure}

Figure~\ref{Fig:lc_different_kappa} demonstrates that lower opacity reduces the photon diffusion time, allowing radiation to escape sooner. With less time trapped in the optically thick, expanding ejecta, adiabatic losses are diminished. Consequently, the light curve rises more rapidly, peaks earlier, and achieves a higher maximum luminosity. The ejecta become optically thin at an earlier epoch, entering the regime where energy deposition and radiative losses balance sooner. In contrast, a higher‐opacity model peaks later, exhibits a lower peak luminosity, and produces a significantly broader light curve.

\subsection{Dependence on Initial Thermal Energy and Progenitor Radius}

The progenitor radius of a supernova can vary significantly, typically ranging from \(\sim 1\) to \(10^3\,R_\odot\). The initial light curve is primarily powered by thermal energy deposited by the outgoing shock wave. A larger pre-explosion radius or greater shock-deposited energy results in a brighter and more prolonged shock-cooling phase. This phase dictates the  behavior of light curve during the first few days post-explosion.  Once radioactive \(^{56}\mathrm{Ni}\) heating dominates, the rise to peak, peak luminosity, and subsequent decline converge and become essentially independent of progenitor radius and shock energy.

\begin{figure*}
    \centering
\includegraphics[width=0.45\textwidth]{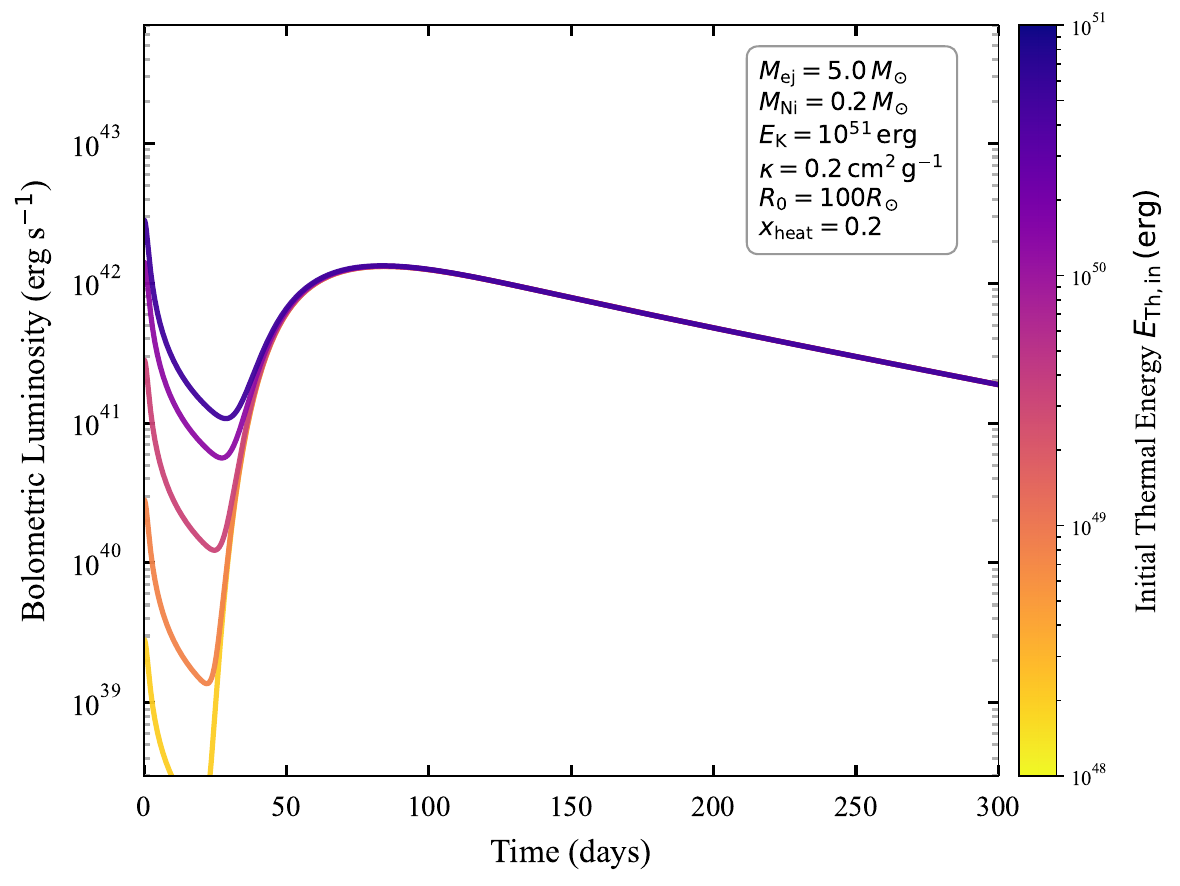}
\includegraphics[width=0.45\textwidth]{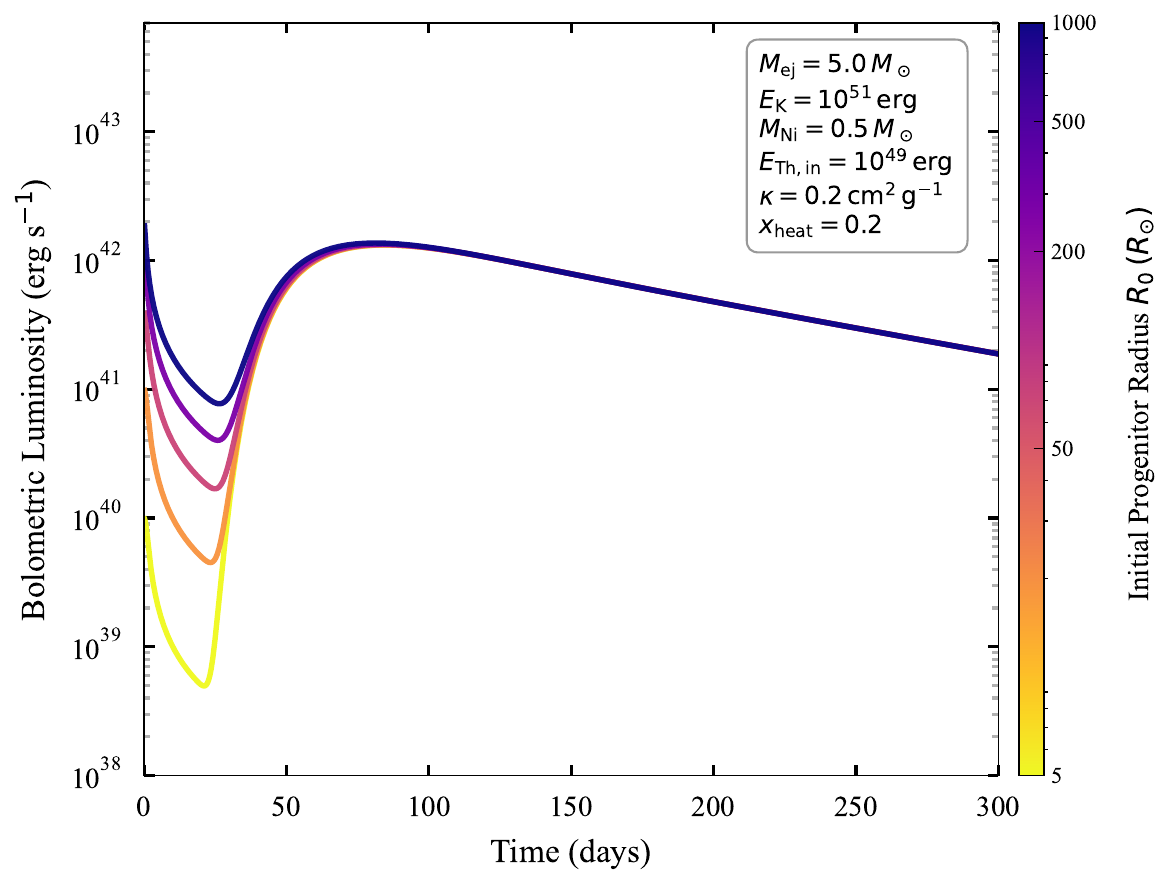}
    \caption{Impact of the 
   the initial thermal energy $E_\mathrm{Th,in}$ and initial progenitor radius $R_{0}$,  on supernova bolometric lightcurves.}
    \label{Fig:lc_different_E_Th_and_R0}
\end{figure*}

Figure \ref{Fig:lc_different_E_Th_and_R0} illustrates how initial thermal energy $E_{\mathrm{Th,in}}$ and initial progenitor radius  $R_{0}$ affect the theoretical bolometric light curves. The left panel demonstrates that a higher initial thermal energy  leads primarily to a brighter early-time luminosity and a higher peak luminosity, with less impact on the decline rate after about 50 days. Conversely, the right panel shows that a larger initial progenitor radius  results in a much slower rise to a broader, potentially slightly lower peak, which occurs later in time compared to supernovae from more compact progenitors.

\subsection{Dependence on Ejecta Density Profile}

\begin{figure*}
    \centering
\includegraphics[width=0.45\textwidth]{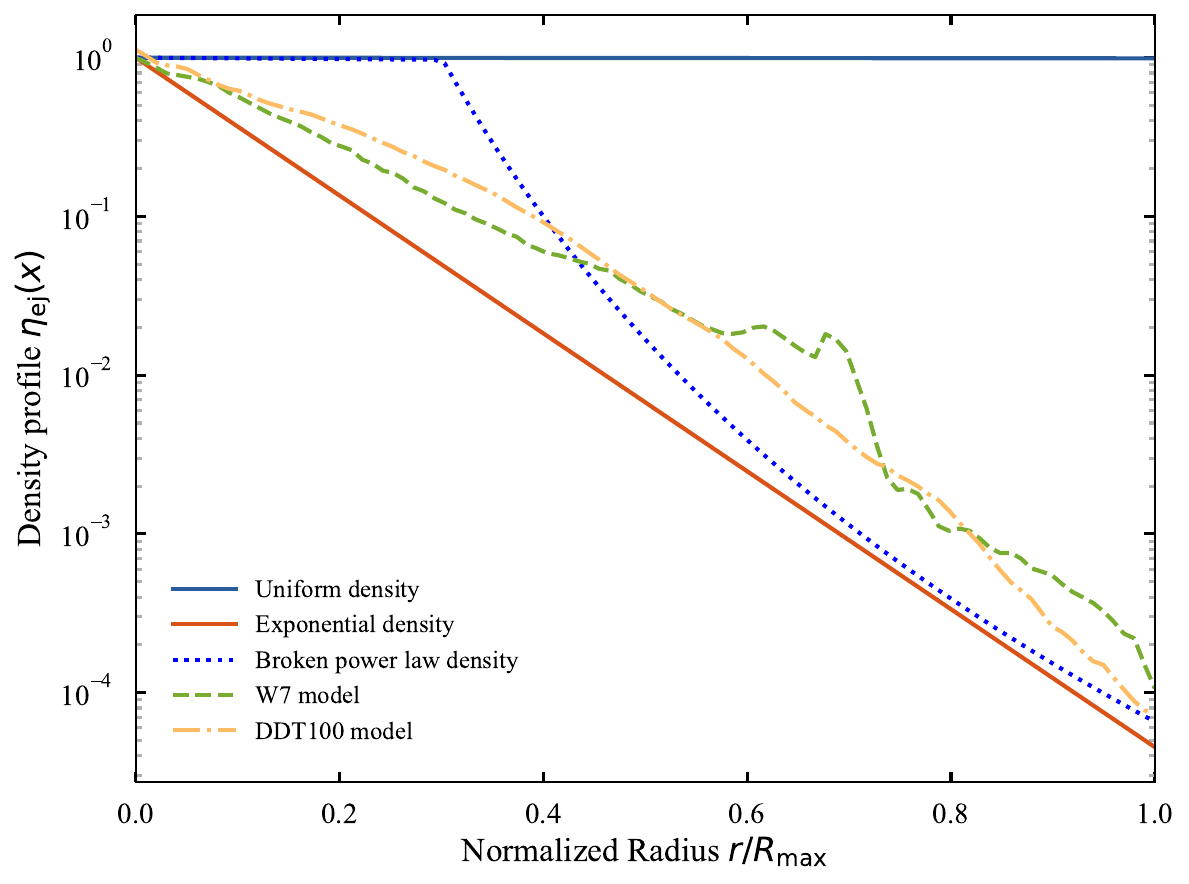}
\includegraphics[width=0.45\textwidth]{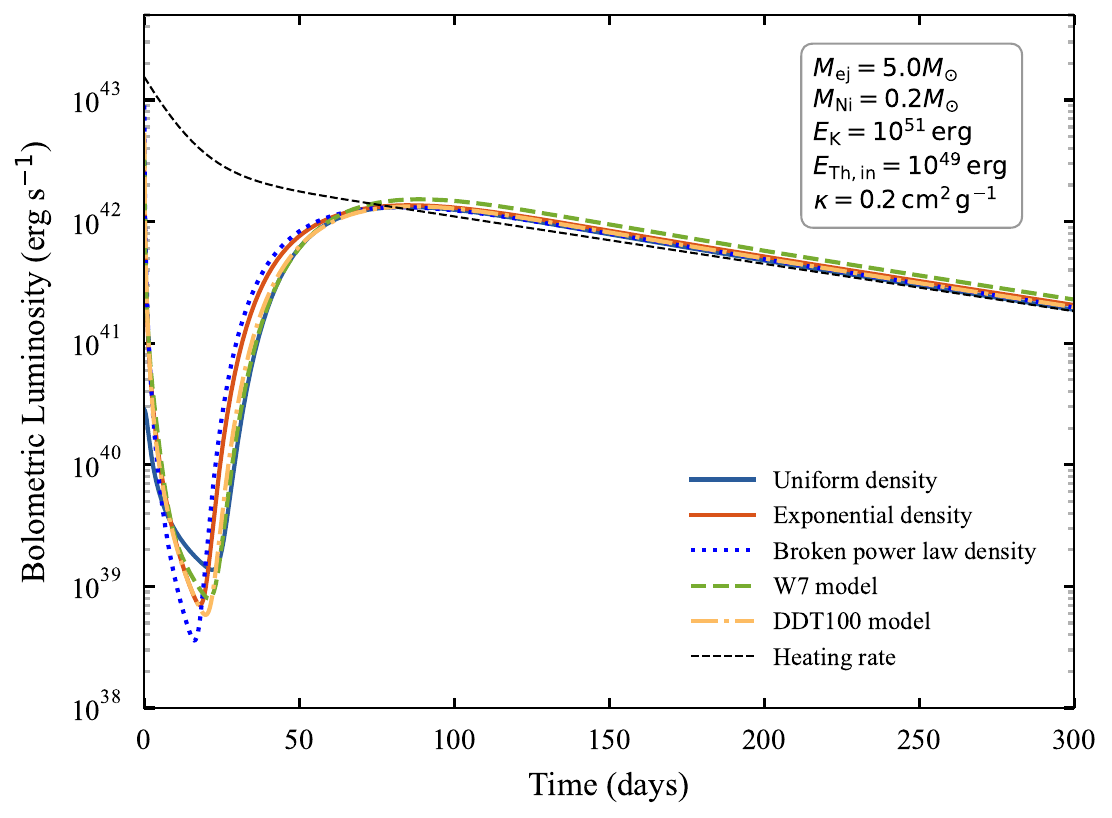}
    \caption{\textit{Left:} Normalized ejecta density profiles $\eta_{\mathrm{ej}}(x)$ versus normalized radius $r/R_{\mathrm{max}}$ for five models: uniform, exponential, broken power law, W7, and DDT100. \textit{Right:} Corresponding bolometric light curves. At default parameters, the peak luminosity and decline rate are nearly independent of the density profile; only the early rise time and late‐time tail show modest variations. The dashed black line indicates the instantaneous radioactive heating rate.}
    \label{Fig:lc_different_Density}
\end{figure*}

Although previous studies have extensively discussed how most of these parameters influence light curves, there has been comparatively little focus on the impact of different density distributions of ejecta. In many past calculations, the ejecta density profile was often assumed to be uniformly distributed. However, numerical simulations of supernova explosion have shown that the actual ejecta density distribution can significantly deviate from uniformity, potentially having a substantial effect on the resulting light curves.  In the canonical W7 deflagration model for Type\,Ia supernovae \citep{Nomoto1984}, once the ejecta attain homologous expansion, the density is well described by an exponential profile with a nearly uniform inner core. In the DDT\,N100 model, angle-averaged profiles from the three-dimensional simulation reveal a nearly uniform core extending to \(v\approx5000\)\,km\,s$^{-1}$, followed by a steep power-law decline \(\rho\propto v^{-7}\) out to \(v\approx15000\)\,km\,s$^{-1}$, and an exponential high-velocity tail containing only a few percent of the total mass \citep{Seitenzahl2013}.

In core-collapse supernova ejecta, once homologous expansion is established, the radial density profile is often modeled as a broken power law: a nearly flat inner core (slope \(\delta\approx0\)–1) extends to a transition velocity \(v_{\rm tr}\), beyond which the density declines steeply with slope \(n\approx9\)–12, reflecting the progenitor’s envelope structure and the shock-accelerated outer layers \citep{Matzner1999}.

The left panel of Figure~\ref{Fig:lc_different_Density} shows that, despite sharing the same total ejecta mass, the five models exhibit markedly different density profiles. As illustrated in the right panel, these structural variations produce nearly identical bolometric light curves: the time of peak luminosity and the post-maximum decline rate are essentially independent of the density profile for the adopted parameters, with only minor discrepancies appearing during the early rise. Consequently, a constant-density ejecta model yields a light curve nearly indistinguishable from one with a more realistic profile, demonstrating that the light-curve shape is largely insensitive to the detailed density structure. This is because the overall mass and energy of the ejecta play a more dominant role in setting the characteristic diffusion timescale, which governs the broad features of the light curve, particularly after peak. Furthermore, the trapping and diffusion of photons in the optically thick ejecta tend to smooth out the effects of smaller-scale density variations, averaging over the detailed structure. The late-time tail of the light curve, powered by radioactive decay, is primarily determined by the total amount of radioactive isotopes synthesized, further reducing the impact of the initial density configuration.

\begin{figure*}[ht!]
    \centering
\includegraphics[width=0.90\textwidth]{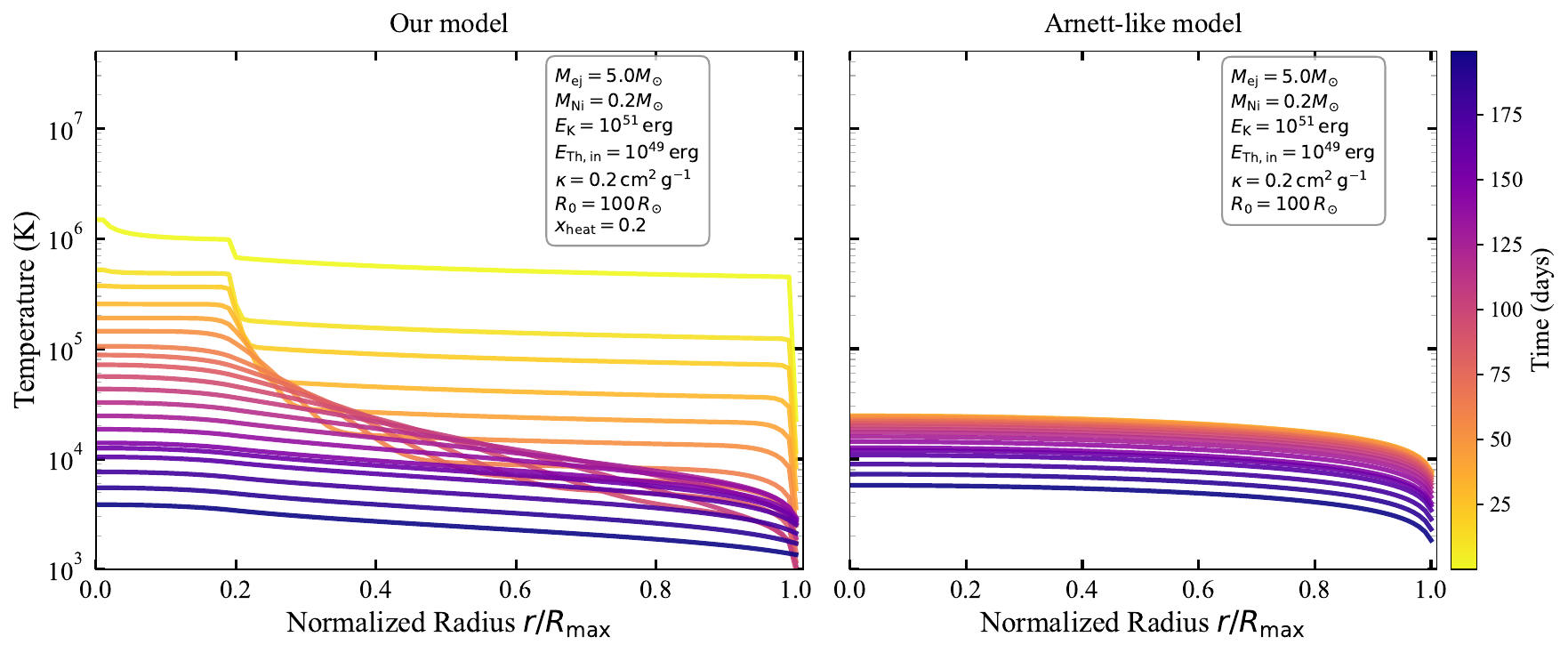}
    \caption{Comparison of the evolution of the internal temperature profile in supernova ejecta between our model (left) and an Arnett‐like model (right). Temperature (K) is plotted versus normalized radius \(r/R_{\max}\), with each curve color‐coded by post‐explosion time (days) as indicated by the color bar.  The input parameters are specified in the legend.
}
    \label{Fig:Temperature_com}
\end{figure*}

\section{Comparison with Previous Analytic Works} \label{Sec:Comparison}

The light‐curve evolution is governed by the energy‐conservation equation, which incorporates heating sources, radiative diffusion, and adiabatic losses. This equation is a nonhomogeneous partial differential equation with nontrivial boundary conditions, precluding a closed‐form solution in general. To extract physical intuition, many authors have developed idealized analytic models that preserve the leading diffusion physics under simplifying assumptions. The two most influential prescriptions are the Arnett model and the one‐zone model, which we briefly summarize below. At the end of this section, we will compare these two models with the spatial temperature evolution and light curves produced by \texttt{TransFit}.

\subsection{Arnett model}
In \cite{Arnett1980}, internal sources of heat (such as the
decay of $^{56}$Ni and $^{56}$Co, pulsar heating, etc.) were
neglected. In this case, the primary source of energy for the light curve is
the residual energy left behind by the rebound shock that sweeps through the
ejecta. Under these assumptions, the energy conservation equation can be
solved using the method of separation of variables. The internal energy
density is assumed to have the form
\begin{equation}
    u (x, t) = u_0 \left( \frac{R_0}{R_{\max}} \right)^4 \Psi (x) \phi (t),
\end{equation}
where $\Psi (x)$ depends only on the spatial variable $x$, and $\phi (t)$
depends only on the time variable $t$.

In this case, the energy conservation equation is homogeneous and can be
written as
\begin{equation}
\frac{1}{x^2 \Psi (x)} \frac{\mathd}{\mathd x} \left[
\frac{x^2}{\eta_{\tmop{ej}} (x)} \frac{\mathd \Psi}{\mathd x} \right] =
\frac{\dot{\phi} t_{\tmop{diff}}}{\phi}  \frac{R_0}{R_{\max}} \equiv - \alpha.
\end{equation}
Since the terms on the left-hand side of this equation depend only on $x$,
while those on the right-hand side depend only on $t$, both must be equal to a
constant independent of either $x$ or $t$. We denote this separation constant by $\alpha$. The spatial part of the equation then reads
\begin{equation} \label{Eq:phi_x}
    \frac{1}{x^2 \Psi (x)} \frac{\mathd}{\mathd x} \left[
\frac{x^2}{\eta_{\tmop{ej}} (x)} \frac{\mathd \Psi}{\mathd x} \right] = -
\alpha.
\end{equation}
The value of $\alpha$ is determined by the boundary conditions. At the center, a reflection boundary condition is applied where the flux
vanishes, or equivalently, $\Psi' (0) = 0$. At the surface, a radiative zero
condition is imposed, which means that the outflow is optically thick, such that $\Psi (1) = 0$ \citep{Pinto2000}. Based on these boundary conditions, for a uniform density distribution $\eta_{\mathrm{ej}} (x) = 1$, the spatial part of the equation gives the eigenvalue:
\begin{equation}
    \alpha_n = n^2 \pi^2, \quad \text{and} \quad \Psi_n = \frac{\sin (n \pi x)}{n
\pi x},
\end{equation}
where $n$ is any positive integer.

\begin{figure*}
    \centering
\includegraphics[width=0.45\textwidth]{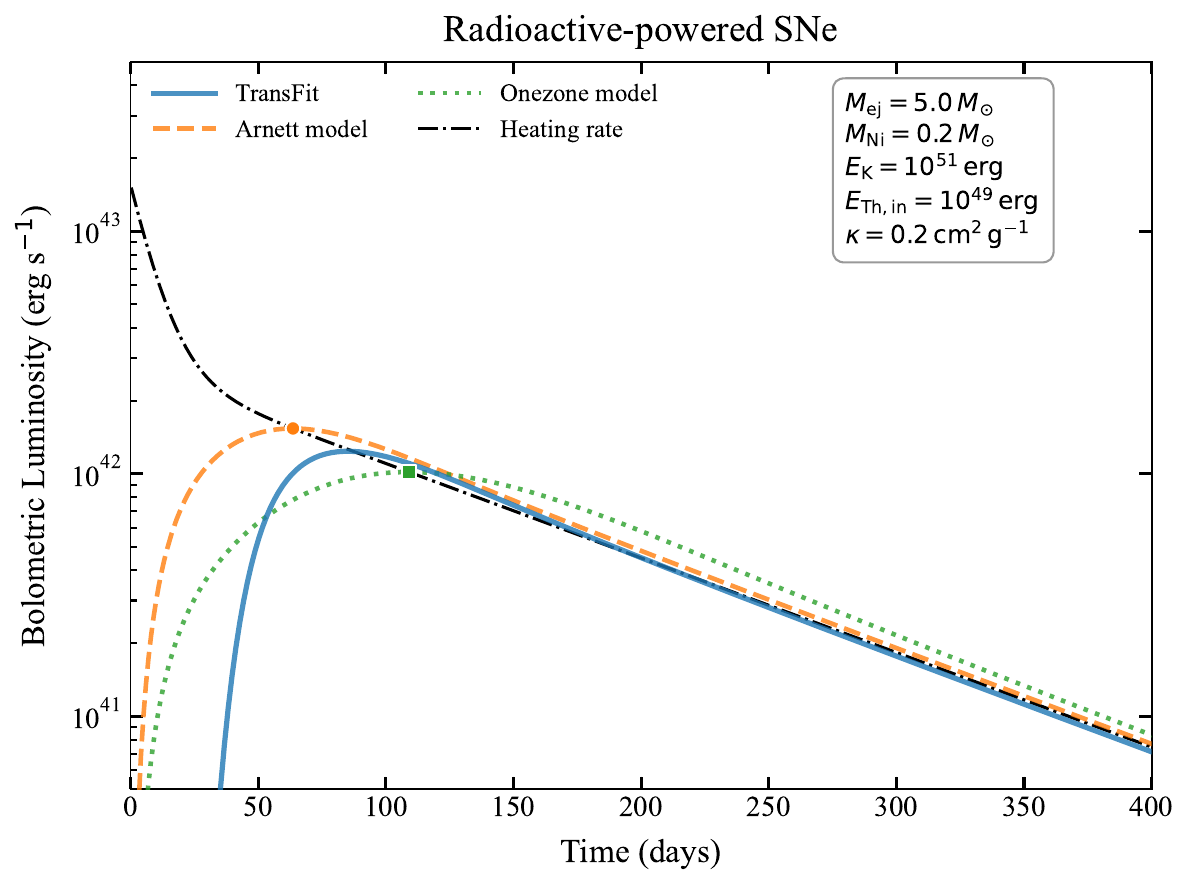}
\includegraphics[width=0.45\textwidth]{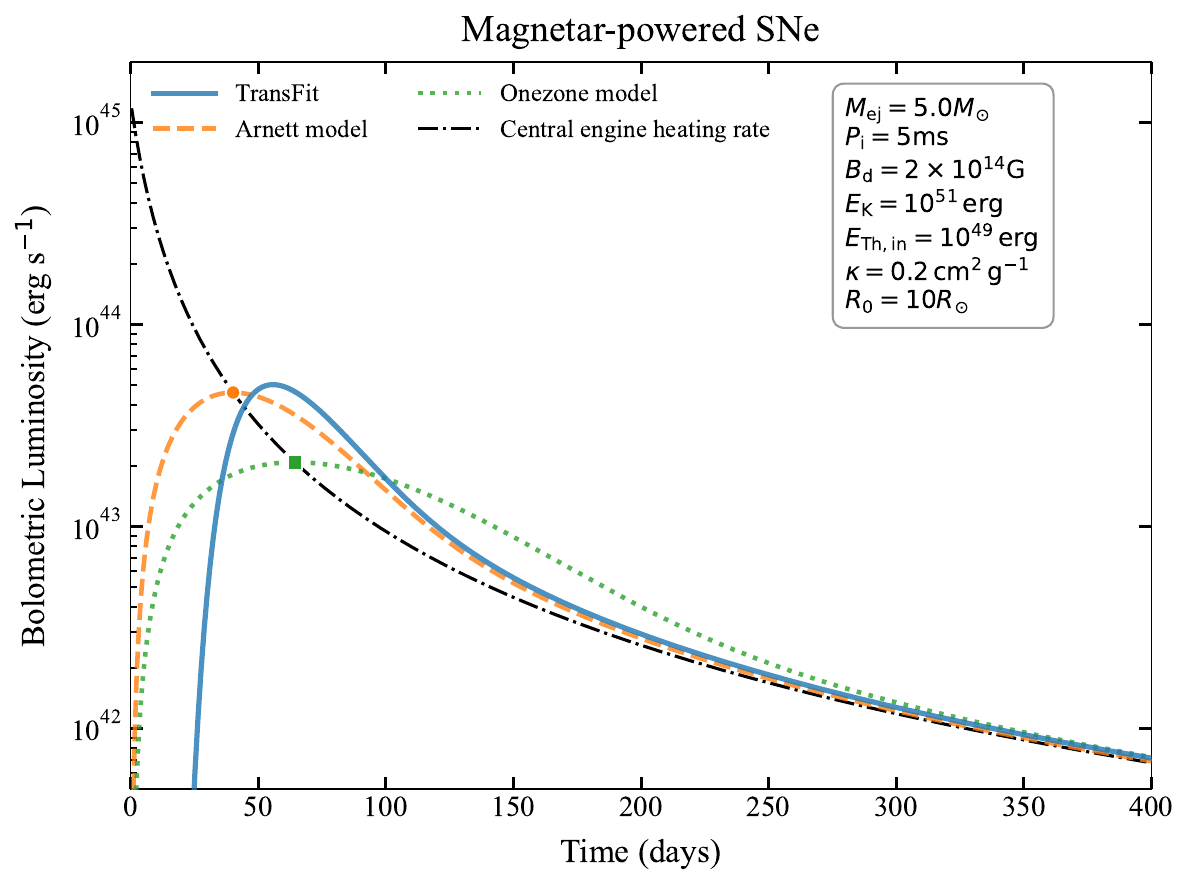}
    \caption{Compares bolometric light-curve predictions from $\texttt{TransFit}$ (solid blue line), the onezone model (dotted green), and the Arnett model (dashed orange). The left panel shows a radioactive-decay–powered scenario, while the right panel illustrates a magnetar-powered scenario. 
}
    \label{Fig:lightcurve_comparision}
\end{figure*}

When \( \eta_{\mathrm{ej}}(x) \) is an arbitrary function, there are no analytic solutions to Eq.~(\ref{Eq:phi_x}), and we must determine the eigenfunctions numerically. The eigenvalues can be calculated using a Rayleigh-Ritz procedure, and a discrete representation of the corresponding eigenfunctions is obtained through a relaxation method \citep{Pinto2000}.

Given the separation constant $\alpha_n$, the temporal part of the solution,
$\phi_n$, is determined by the homogeneous equation:
\begin{equation}
    \frac{\mathrm{d} \phi_n}{\mathrm{d} y} = - \alpha_n  \left( 1 +
\frac{y}{y_{\mathrm{ex}}} \right) \phi_n,
\end{equation}
where $y = \frac{t}{t_{\mathrm{diff}}}$ and $y_{\mathrm{ex}} =
\frac{t_{\mathrm{ex}}}{t_{\mathrm{diff}}}$.  This expression is equivalent to the one in \cite{Arnett1980}. Without loss of
generality, we can take $\phi (0) = 1$, yielding the solution:
\begin{equation}
    \phi_n (y) = \exp \left[ - \alpha_n  \left( y + \frac{y^2}{2 y_{\mathrm{ex}}}
\right) \right].
\end{equation}

The general solution is typically a sum of modes (an infinite
series) for the possible values of $\alpha$:
\begin{equation}
    u (x, t) = u_0 \left( \frac{R_0}{R_{\max}} \right)^4  \sum_{n = 1}^{\infty}
\phi_n (t) \Psi_n (x).
\end{equation}
However, in \cite{Arnett1980,Arnett1982}, only the uniform density distribution was considered, and only the first eigenvalue, $\alpha = \pi^2$, was taken \footnote{\cite{Pinto2000} demonstrated that a complete solution for the energy density requires representation as an infinite series of higher-order eigenfunctions, whose normalization is determined by the spatial distribution of the heating source and the boundary conditions. However, despite its theoretical thoroughness, this formulation involving higher-order eigenmodes is few used in practical analyses due to its inherent complexity.}. Under this approximation, a single eigenmode describes the shape of the internal
energy density, which does not change over time. In this case, the solution
simplifies to:
\begin{equation}
    u (x, t) = u_0 \left( \frac{R_0}{R_{\max}} \right)^4 \frac{\sin (\pi x)}{x}
\exp \left[ - \pi^2  \left( y + \frac{y^2}{2 y_{\mathrm{ex}}} \right) \right]
.
\end{equation}
In this case, the bolometric luminosity is determined by the flux at the surface of the ejecta. The luminosity is given by
\begin{equation}
    L (t) = \frac{E_{\mathrm{Th}, \mathrm{in}}}{t_{\mathrm{diff}}}
\exp \left[ - \pi^2  \left( \frac{t}{t_{\mathrm{diff}}} + \frac{t^2}{2
t_{\mathrm{diff}} t_{\mathrm{ex}}} \right) \right].
\end{equation}
When $t > 2 R_0 / v_{\max}$, the term $t^2$ becomes dominant, and the
luminosity decays as $L_{\mathrm{bol}} \propto \exp \left( - \frac{t^2}{\tau_{\mathrm{m}}^2}
\right),$ which falls off as a Gaussian rather than an exponential in time. The characteristic diffusion timescale is defined as
\begin{equation}
    \tau_{\mathrm{m}} \equiv \sqrt{\frac{2 t_{\mathrm{diff}}
t_{\mathrm{ex}}}{\pi^2}} = \sqrt{\frac{2 \kappa M_{\mathrm{ej}}}{\beta
cv_{\max}}},
\end{equation}
which represents the geometric mean of the diffusion timescale and the
expansion timescale, where $\beta \equiv \frac{4 \pi \alpha I_{\mathrm{M}}}{3}
\approx 13.8$ is a constant derived from the integration of the density
profile of the expanding ejecta.

\cite{Arnett1982} considered the heating effect of  $^{56}$Ni in the ejecta. In this scenario, the fundamental equation can be expressed as:
\begin{equation}\label{Eq:A82}
\begin{split}
\frac{R_0}{R_{\max}} \frac{1}{\phi} \frac{\mathrm{d} \phi}{\mathrm{d} y} &= \frac{1}{x^2 \Psi (x)} \frac{\mathrm{d}}{\mathrm{d} x} \left[ \frac{x^2}{\eta_{\mathrm{ej}} (x)} \frac{\mathrm{d} \Psi}{\mathrm{d} x} \right] \\
&\quad + \frac{\xi_{\mathrm{heat}} (x) \eta_{\mathrm{ej}} (x)}{\Psi (x)} \frac{f_{\mathrm{heat}} (y)}{\phi},
\end{split}
\end{equation}
where the second term on the right-hand side represents the heating source. Here, we adopt $u_0 = t_{\mathrm{diff}}\,\rho_0\,\epsilon_{\mathrm{heat},0}$ to simplify the equation to the form above. This equation is non-homogeneous and, in general, cannot be solved by separation of variables. \citet{Arnett1982} therefore assumed
\begin{equation}
b \equiv\frac{\xi_{\mathrm{heat}}(x)\,\eta_{\mathrm{ej}}(x)}{\Psi(x)}
=\text{constant}, \quad\text{for all }x,
\end{equation}
which renders the equation separable.  Only under this assumption can the equation be solved by separation of variables. However, this assumption implies that the heating source has no impact on the spatial distribution of internal energy, which is clearly non-physical. Specifically, when the heating source is centrally concentrated, with limited distribution in the outer regions, energy can still be transferred to these areas through diffusion. The spatial distribution of the internal energy determines that the equation
remains in the same form as Eq.~(\ref{Eq:phi_x}).  The time-dependent part of Eq.~(\ref{Eq:A82}) can be written as
\begin{equation}
    \frac{\mathrm{d} \phi}{\mathrm{d} y} + \alpha \left( 1 + \frac{y}{y_{\mathrm{ex}}} \right) \phi - b\left( 1 + \frac{y}{y_{\mathrm{ex}}} \right) f_{\mathrm{heat}}(y) = 0,
\end{equation}
which can be solved by the integrating‐factor method, as detailed in Appendix~\ref{Sec:Integrating_factor}.  The resulting semi‐analytic solution is
\begin{equation}
\begin{split}
\phi (y) &= b e^{- \alpha \left( y + \frac{y^2}{2 y_{\mathrm{ex}}} \right)} \int_{0}^{y} e^{\alpha \left( y' + \frac{y'^2}{2 y_{\mathrm{ex}}} \right)}\left( 1 + \frac{y'}{y_{\mathrm{ex}}} \right) f_{\mathrm{heat}} (y') \mathrm{d} y' \\
&\quad + C e^{- \alpha \left( y + \frac{y^2}{2 y_{\mathrm{ex}}} \right)}.
\end{split}
\end{equation}
Imposing the initial condition \(\phi(0)=1\) gives the constant of integration $C=1$. 

Under the single-mode approximation ($\alpha = \pi^2$), the bolometric luminosity reduces to:
\begin{equation}
\begin{split}
L(t) &= \frac{2 L_{\mathrm{heat} 0}}{\tau_{\mathrm{m}}} e^{- \left( \frac{t^2 + 2 t_{\mathrm{ex}} t}{\tau_{\mathrm{m}}^2} \right)} \int_0^t \left( \frac{t_{\mathrm{ex}} + t'}{\tau_{\mathrm{m}}} \right) e^{\left( \frac{{t'}^2 + 2 t_{\mathrm{ex}} t'}{\tau_{\mathrm{m}}^2} \right)} f_{\mathrm{heat}} (t') \mathrm{d} t' \\
&\quad + L_{\mathrm{Th} 0} e^{- \left( \frac{t^2 + 2 t_{\mathrm{ex}} t}{\tau_{\mathrm{m}}^2} \right)},
\end{split}
\end{equation}
where the first term on the right-hand side represents the contribution to the bolometric luminosity from a time-dependent heating source described by $L_{\mathrm{heat}} (t) = L_{\mathrm{heat} 0} f_{\mathrm{heat}} (t)$. The second term on the right-hand side represents the cooling luminosity arising from the initial internal energy, where $L_{\mathrm{Th} 0} \equiv E_{\mathrm{Th,in}}/t_{\mathrm{diff}}$ is the characteristic cooling luminosity.

For small initial radius case, the terms related to $t_{\mathrm{ex}}$ can be neglected. Additionally, neglecting the cooling contribution from the initial internal energy, the bolometric light curve can be expressed in a more widely used form:
\begin{equation} \label{Eq:L_bol_Arnett}
    L (t) = \frac{2 L_{\tmop{heat} 0}}{\tau_{\text{m}}^2} e^{-
\left( \frac{t^2}{\tau_{\text{m}}^2} \right)} \times \int_0^t t' e^{\left(
\frac{{t'}^2}{\tau_{\text{m}}^2} \right)} f_{\tmop{heat}} (t') \mathd t'.
\end{equation}
Here, $\tau_{\mathrm{m}}$ represents the characteristic diffusion time through the ejecta with mass $M_{\mathrm{ej}}$, and is typically approximated as the rise timescale of the light curve \citep{Nicholl2015}.

\subsection{onezone model}

Another widely used semi-analytic framework is the onezone model \citep{Arnett1979,Kasen2010,Yu2015}. It treats the ejecta as a single, homogeneous sphere with average density, opacity, temperature, and internal energy, expanding at a characteristic velocity $v_{\tmop{ej}} = \sqrt{2 E_{\text{K}} / M_{\tmop{ej}}}$. The evolution of the total internal energy, $E_{\mathrm{int}}$, within the ejecta is governed by the first law of thermodynamics:
\begin{equation} \label{Eq:Basic_onezone}
    \frac{\mathd E_{\tmop{int}}}{\mathd t} = - P \frac{\mathd V}{\mathd t} - L +
L_{\tmop{heat}} (t),
\end{equation}
where $V$ is the ejecta volume, $L_{\tmop{heat}} (t)$ is the total input heating rate. $L$ radiative luminosity at the ejecta surface, is approximated from the diffusion equation
\begin{equation}
\frac{L}{4\pi R^2}=\frac{c}{3\kappa\rho}\frac{\partial E_\mathrm{int}/V}{\partial r}\approx\frac{c}{3\kappa\rho}\frac{E_\mathrm{int}/V}{R}.\end{equation}
The final approximation assumes a linear spatial gradient of the internal energy within the ejecta, implying that the spatial distribution of the internal energy profile maintains a self-similar form. In this case, the emergent luminosity \( L \) and the total internal energy \( E_{\mathrm{int}} \) are related by the following equation:
\begin{equation}
    L=\frac{E_\mathrm{int}t}{\tau_\mathrm{d}^2},
\end{equation}
where $\tau_{\text{d}}$ is the effective diffusion
time in one-zone model,  given by
\begin{equation}
\tau_\mathrm{d}\equiv\left(\frac{3\kappa M_\mathrm{ej}}{4\pi v_\mathrm{ej}c}\right)^{1/2}.
\end{equation}

Rearranging the Eq.(\ref{Eq:Basic_onezone}) gives 
\begin{equation}
    \frac{\mathd L}{\mathd t} = \frac{t}{\tau_{\text{d}}^2} (L_{\tmop{heat}} -L),
\end{equation}
this ordinary differential equation can also solved by the integrating-factor method.  The solution is obtained as 
\begin{equation}
L (t) = e^{- \frac{t^2}{2 \tau_{\mathrm{d}}^2}} \int_0^t \frac{t'}{\tau_{\mathrm{d}}^2} L_{\mathrm{heat}} (t') e^{\frac{{t'}^2}{2 \tau_{\mathrm{d}}^2}} \mathrm{d} t'.
\end{equation}

\subsection{Light curve from different models}
Both the one-zone model and the separation of variables approach yield the same expression for the light curve, with the only difference being the values of the diffusion timescale. The use of \( \tau_{\text{m}} \) and \( \tau_{\text{d}} \) as characteristic diffusion timescales in the semi-analytic model is based on the assumption of a self-similar internal energy density profile (i.e., the spatial distribution of internal energy remains unchanged over time). Under this assumption, a natural consequence is Arnett's law, which states that
\begin{equation}
  L_{\mathrm{peak}} = L_{\mathrm{heat}}(t_{\mathrm{peak}}), 
\end{equation}
where \( L_{\mathrm{peak}} \) is the luminosity at the time of the peak \( t_{\mathrm{peak}} \), and \( L_{\mathrm{heat}}(t_{\mathrm{peak}}) \) is the heating rate at that same time.

Arnett's law is a fundamental concept in supernova physics that relates the peak luminosity of a supernova light curve to the instantaneous rate of heating at that time. It is widely used to infer parameters of the heating source, such as \( M_{\mathrm{Ni}} \) \citep{Stritzinger2006}. However, Arnett’s law does not hold universally, and its accuracy depends on the spatial distribution of the heating source. For centrally concentrated heating sources, Arnett’s law tends to underestimate the true peak luminosity, with the error being systematically larger for more centralized heating distributions \citep{Khatami2019}. 

In reality, the time-dependent propagation of a diffusion wave causes the luminosity to deviate from a strict proportionality to internal energy, i.e., \( L \propto E_{\mathrm{int}} t \). The spatial distribution of the heating source affects the evolution of the internal energy profile (as shown in the right panel of Figure \ref{Fig:Temperature_com}). Therefore, in general, it is not possible to describe how photons diffuse out of the ejecta using a single characteristic diffusion timescale, $\tau_{\mathrm{m}}$ or $\tau_{\mathrm{d}}$. Such an approach can only be applied under the assumption that the internal energy profile remains unchanged, and Arnett’s law only holds under this premise.

\begin{figure*}
    \centering
\includegraphics[width=0.45\textwidth]{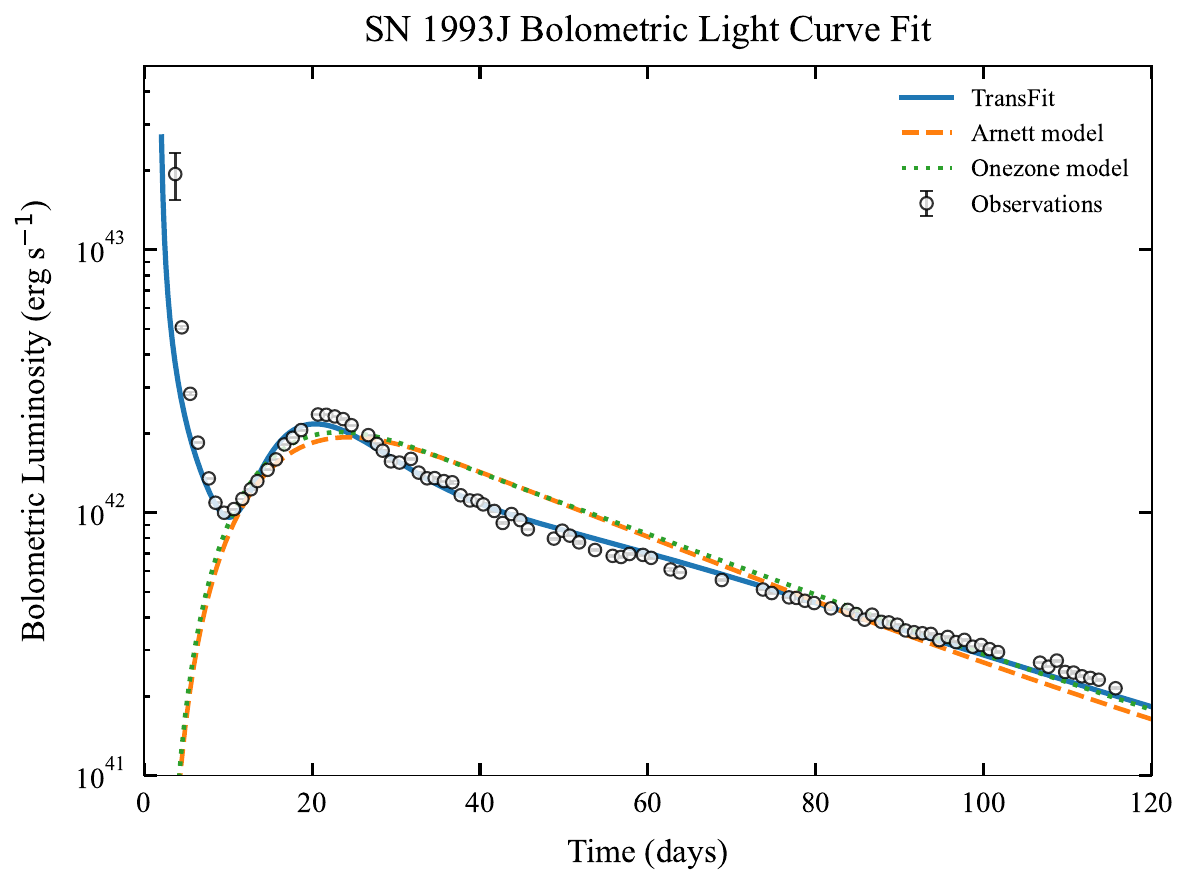}
\includegraphics[width=0.45\textwidth]{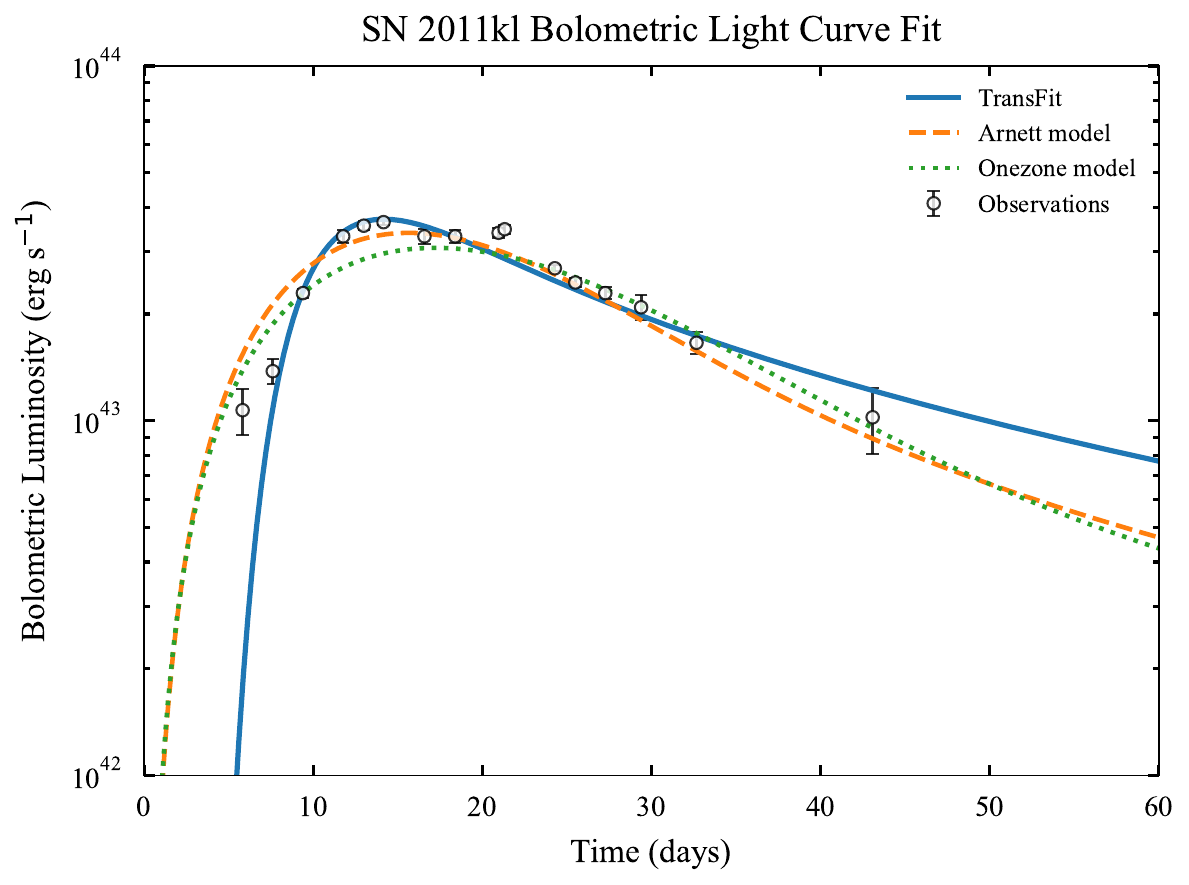}
    \caption{Bolometric light‐curve fits for SN 1993J (left) and SN 2011kl (right). Observations (circles with error bars) are compared to three theoretical models: \texttt{TransFit} (solid blue), the Arnett model (dashed orange), and the one‐zone model (dotted green). Bolometric data for SN 1993J are from \citet{Richmond1994}, and data for SN 2011kl are from \citet{Kann2019}.
}
    \label{Fig:SN1993j}
\end{figure*}

Figure~\ref{Fig:lightcurve_comparision} compares bolometric light curves computed with three radiation-transport models. The \texttt{TransFit} solutions remain essentially dark for the first \(\sim10\)–\(20\)\,days—the photon-diffusion time required to traverse the optically thick ejecta—where the duration depends on \(M_{\rm ej}\), the mean opacity \(\kappa\), and the radial heating profile \citep{Piro2013,Khatami2019}. In contrast, analytic models (Arnett and one-zone) assume instantaneous, homogeneous energy deposition, yielding a steep rise from \(t=0\) and earlier, marginally brighter peaks. After diffusion breakout, all three methods converge to similar peak luminosities; the onezone model’s peak is broader, as its gray-opacity approximation overestimates the diffusion time. In both radioactive- and magnetar-powered scenarios, the post-maximum decline tracks the instantaneous heating rate, with slight offsets due to differences in \(\gamma\)-ray trapping (radioactive) or spin-down energy thermalization (magnetar).

\begin{deluxetable}{lccccc}
\tablecaption{Model parameters for SN\,1993J.\label{tab:sn1993j_params}}
\tablehead{
\colhead{Model} &
\colhead{$M_{\rm ej}$} &
\colhead{$M_{\rm Ni}$} &
\colhead{$E_{\rm K}$} &
\colhead{$R_{\rm 0 }$}&
\colhead{$E_{\rm Th,in}$}\\[-4pt]
\colhead{}&       
\colhead{($M_\odot$)} &
\colhead{($M_\odot$)} &
\colhead{($10^{51}$ erg)} &
\colhead{($10^{13}$ cm)}&
\colhead{($10^{49}$ erg)}
}
\startdata
W94$^{1}$            & 1.4  & 0.07 & 1.3 & 4.0 & -- \\
N16$^{2}$            & 2.3  & 0.10 & 2.4 & 3.0 & -- \\
Arnett         & 3.0  & 0.12 & 2.5 & --  & -- \\
Onezone         & 2.0  & 0.12 & 2.5 & --  & -- \\
\texttt{TransFit}    & 1.8  & 0.10 & 1.5 & 3.1 & 4.0 \\
\enddata
\vspace{0.5em}
\textbf{References.}(1)\,\cite{Woosley1994}, (2)\,\cite{Nagy2016}.
\end{deluxetable}

\section{Apply TransFit for SN 1993J and SN 2011kl} \label{Sec:Fitting}

In this section, we apply \texttt{TransFit} to two well-observed supernovae: SN\,1993J as a prototypical radioactive-decay–powered event, and SN\,2011kl as a canonical magnetar-powered transient. For each case, we compile published bolometric light curves and adopt distance and extinction estimates from the literature before fitting these data with identical priors on key physical parameters—ejecta mass, kinetic energy, opacity, and heating source. Although extending this analysis to a larger supernova sample is beyond the scope of this work, we plan to undertake such an effort in future studies.

\subsection{SN  1993J}
SN\,1993J, discovered on 1993 March 28 in the nearby spiral galaxy NGC\,3031 (M81), is a prototypical Type IIb supernova with exceptionally good early–time photometric coverage. Its progenitor and explosion parameters have been derived via a variety of light‐curve modeling studies\citep{Shigeyama1994,Woosley1994,Blinnikov1998,Nagy2016}.

The left panel of Figure~\ref{Fig:SN1993j} compares the observed bolometric light curve of SN\,1993J  with different  theoretical models calibrated to the parameters in Table~\ref{tab:sn1993j_params}. The \texttt{TransFit} solution, which adopts the lowest ejecta mass \(M_{\mathrm{ej}}\approx1.8\,M_\odot\) and an initial thermal reservoir \(E_{\mathrm{Th,in}}=4\times10^{49}\)\,erg, accurately reproduces both the brief shock-cooling dip at \(t\lesssim10\)\,days and the first-peak luminosity at \(t\approx22\)\,days. In contrast, the Arnett model  and the one-zone model —both tuned to larger \(M_{\mathrm{ej}}\) and \(E_{\rm K}\)—match the post-peak decline but underpredict the depth of the early dip and slightly overshoot the peak. The systematic offset during the initial cooling phase underscores the importance of explicitly modeling shock-heated surface layers, while the convergence of all three curves at \(t\gtrsim40\)\,days reflects the dominance of the radioactive \(^{56}\mathrm{Co}\) decay tail. Overall, the inclusion of a finite initial thermal energy and a modestly lower ejecta mass—both consistent with hydrodynamic studies—yields a superior, simultaneous fit to SN\,1993J’s double-peaked light curve.

\begin{deluxetable}{lccccc}
\tablecaption{Model parameters for SN\,2011kl.}
\label{tab:sn2011kl_params}
\tablehead{
\colhead{Model} &
\colhead{$M_{\rm ej}$} &
\colhead{$P_{\rm i}$} &
\colhead{$B_{\rm d}$} &
\colhead{$E_{\rm K}$} \\[-4pt]
\colhead{}&  
\colhead{($M_\odot$)} &
\colhead{(ms)} &
\colhead{($10^{14}$ G)} &
\colhead{($10^{51}$ erg)}
}
\startdata
    G15$^{1}$     & 2.4  & 12.2 & 7.5 & 5.5 \\
    C16$^{2}$      & 5.2  & 11.0 & 13.0 & 22.8 \\
    W17$^{3}$     & 3.1  & 10.9 & 6.6 & 13.8 \\
    Arnett   & 2.0  & 10.6 & 8.0 & 2.0 \\
    Onezone   & 2.0  & 9.6  & 9.0 & 2.0 \\
    \texttt{TransFit} & 1.4  & 10.8 & 5.0 & 5.0 \\
\enddata
\vspace{0.5em}
\textbf{References.}(1)\,\cite{Greiner2015}, (2)\,\cite{Cano2016}, (3)\,\cite{Wang2017}.
\end{deluxetable}

\subsection{SN 2011kl}

SN\,2011kl is an exceptionally luminous, rapidly evolving Type Ic-BL supernova that accompanied the ultra–long gamma-ray burst GRB\,111209A at redshift \(z=0.677\). With a peak bolometric luminosity exceeding \(3\times10^{44}\)\,erg\,s\(^{-1}\) and a blue, featureless continuum spectrum, SN\,2011kl occupies the transitional regime between classical GRB–associated supernovae and SLSNe \citep{Greiner2015}.

Several studies have modeled the light curve of SN\,2011kl. The bolometric ligtcurve cannot be reproduced with a pure \(^{56}\mathrm{Ni}\) decay model but is well fitted by a magnetar engine \citep{Greiner2015,Cano2016,Wang2017}, as well as by hybrid magnetar plus \(^{56}\mathrm{Ni}\) models \citep{Metzger2015,Bersten2016}. Alternative explanations include a white-dwarf tidal-disruption event \citep{Ioka2016} and a collapsar scenario involving a stellar-mass black hole and a fallback accretion disk \citep{Gao2016}.

The right panel of Figure~\ref{Fig:SN1993j} contrasts the observed bolometric light curve of the ultra-luminous Type Ic-BL SN\,2011kl with different magnetar models using the parameters in Table~\ref{tab:sn2011kl_params}. The \texttt{TransFit} solution exhibits a ``dark phase,” remaining essentially undetectable for the first \(\sim10\)\,days—a diffusion delay set by its relatively low ejecta mass (\(M_{\mathrm{ej}}=1.4\,M_\odot\)), moderate opacity, and centralized heating profile. After diffusion breakout, the \texttt{TransFit} curve rises steeply, reproducing both the peak luminosity (\(\sim3\times10^{43}\)\,erg\,s\(^{-1}\) at \(t\approx13\) days) and the subsequent decline with high fidelity. By contrast, the Arnett model  and the one-zone model both configured with \(P_{\rm i}=10.8\)\,ms and \(B_{\rm d}\approx(8-9)\times10^{14}\)\,G—rise immediately from \(t=0\) due to their assumption of instantaneous energy mixing, causing them to overpredict the early luminosity and underestimate the time of peak. All three models converge after \(\sim30\)\,days, reflecting the decreasing influence of diffusion delays once the magnetar luminosity falls below the diffusion limit.

\begin{figure*}
    \centering
\includegraphics[width=0.85\textwidth]{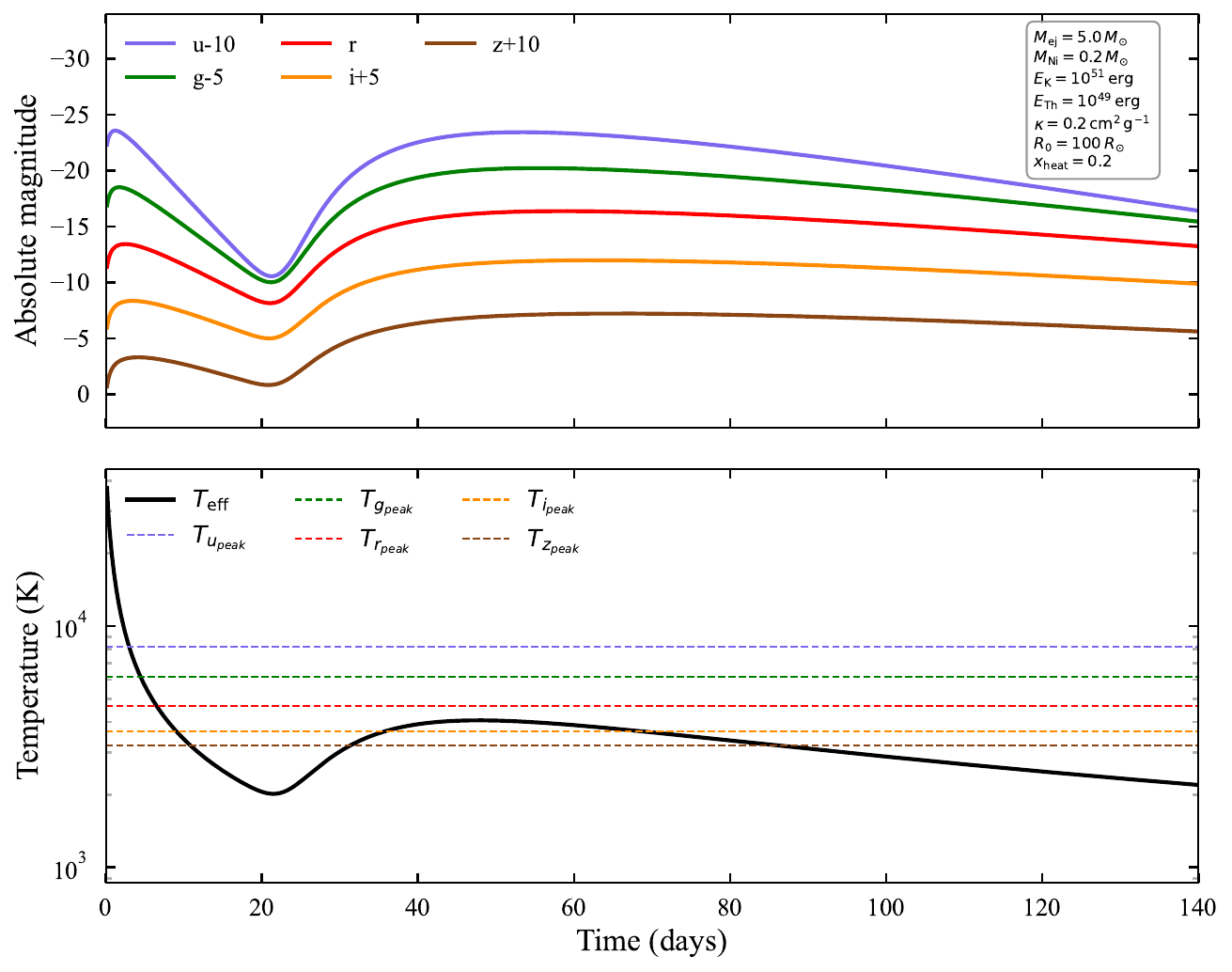}
\caption{Theoretical light-curve and temperature evolution. \textbf{Top:} Absolute magnitudes in the \(u\), \(g\), \(r\), \(i\), and \(z\) bands (offset vertically for clarity) as a function of time. \textbf{Bottom:} Evolution of the effective temperature \(T_{\rm eff}\) (solid black line) alongside the band-specific temperatures at peak luminosity (dashed horizontal lines).
}
    \label{Fig:lightcurve_mag}
\end{figure*}

\section{Discussion and Conclusions} \label{Sec:Conclusion}

We have introduced \texttt{TransFit}, a fast, finite-difference framework that solves the time‐dependent energy‐conservation equation—including radiative diffusion, centrally or radioactively driven heating, and homologous expansion—in a fraction of a second per model. By combining the speed of analytic prescriptions with the fidelity of radiation‐hydrodynamics simulations, \texttt{TransFit} enables rapid, accurate light‐curve fitting.

A major strength of \texttt{TransFit} is its explicit treatment of time‐dependent radiative diffusion and nonuniform heating, which exposes the limitations of Arnett’s Law. Arnett’s Law equates the peak luminosity to the instantaneous heating rate under the assumption of a self‐similar, unchanging internal energy profile. Our results show that when heating is centrally concentrated or spatially stratified, this assumption breaks down and Arnett’s Law systematically underestimates peak luminosities. By modeling the diffusion wave’s propagation through an evolving energy profile, \texttt{TransFit} demonstrates that luminosity deviates from simple proportionality to instantaneous heating. Consequently, \texttt{TransFit} does not depend on a single diffusion timescale—delivers more accurate light‐curve predictions especially in cases where the foundational assumptions of Arnett's Law are violated.

Systematic exploration of ejecta mass, kinetic energy, opacity, initial thermal reservoir, progenitor radius, and \(^{56}\mathrm{Ni}\) mixing reproduces the qualitative trends seen in detailed simulations: broader, fainter peaks for more massive ejecta; faster, brighter peaks for lower opacity; and the transition from shock‐cooling to radioactive or magnetar power.

Simultaneous fits to the double‐peaked Type\,IIb SN\,1993J and the luminous magnetar‐powered SN\,2011kl demonstrate that a single, physically motivated parameter set can reproduce the full bolometric evolution—including the early shock‐cooling phase and the post‐peak decline—using lower ejecta masses and more realistic heating geometries than traditional analytic models.

The Einstein Probe (EP) mission has uncovered a new class of supernovae preceded by fast X-ray transients (FXTs; e.g., EP~240414a, EP~241021a, EP~250108a) without accompanying gamma-ray triggers \citep{Sun2024,Li2025,Dalen2025,Rastinejad2025}. In its first year, EP detected 12 high–signal‐to‐noise FXTs and identified optical‐counterpart candidates for five of them \citep{Aryan2025}. Owing to EP’s wide field of view and rapid X-ray alerts, we can initiate optical follow-up within hours of the trigger, capturing the ensuing supernova during its shock-cooling phase. These early observations allow us to track the luminosity evolution immediately after shock breakout and to place tight constraints on the progenitor radius, ejecta mass, and initial thermal energy.

\texttt{TransFit} provides a self-consistent modeling of the transition from shock-cooling to \(^{56}\mathrm{Ni}\)-powered light curves. As shown in Figure~\ref{Fig:lightcurve_mag}, during the initial ``shock-cooling'' phase (the first \(\sim 10\)--20 days), the bluer bands (e.g., \textit{u} and \textit{g}) exhibit a stronger early peak and a deeper dip than the redder bands (\textit{r}, \textit{i}, \textit{z}), reflecting the rapid decline in photospheric temperature immediately after shock breakout. As the ejecta cools, the \textit{u}-band fades fastest, while the redder bands decline more slowly and recover sooner. This wavelength-dependent behavior directly follows the evolution of the effective temperature \(T_{\mathrm{eff}}\) (solid black line in the bottom panel): when \(T_{\mathrm{eff}}\) falls below each band’s characteristic peak-sensitivity temperature (dashed horizontal lines), that band switches from brightening to fading. Hence, the ordering and timing of the band-specific maxima encode the cooling history of the ejecta during the shock-cooling phase and seamlessly connect to the subsequent \(^{56}\mathrm{Ni}\)-dominated rise to peak.

The interaction of supernova ejecta with a surrounding circumstellar medium (CSM) generates a strong shock that converts kinetic energy into observable radiation. Shock heating can therefore power the light curves of some supernovae. Type IIn events show clear spectral signatures of CSM interaction narrow hydrogen  emission lines indicate a shock propagating into dense circumstellar material and depositing kinetic energy \citep{Chevalier1982,Chevalier1994}. Depending on the CSM density profile and structure, interaction-powered transients can exhibit a wide variety of light-curve morphologies \citep{Chatzopoulos2012,Moriya2013,Liu2018b,Khatami2024}. In this scenario, the energy source propagates outward with the shock rather than being centrally concentrated. We will present a dedicated treatment of CSM interaction in the next installment of the \texttt{TransFit} series—\texttt{TransFit-CSM} (Zhang et al.\ 2025, in preparation).

With its fast execution, \texttt{TransFit} can survey extensive parameter spaces within hours on a standard desktop computer, empowering Bayesian inference for the thousands of transients that next-generation time-domain surveys (e.g., ZTF, Mephisto,  WFST, LSST, CSST) are anticipated to deliver. Future work includes extending its capabilities to incorporate wavelength-dependent opacities and line-blocking effects; coupling with spectral-synthesis codes for multi-band fitting; incorporating gamma-ray and X-ray leakage at late times; and relaxing the assumption of spherical symmetry to treat aspherical outflows and viewing-angle effects. These upgrades aim to cement \texttt{TransFit} as a key community tool for the real-time interpretation of the rapidly growing time-domain dataset.

\begin{acknowledgements}
We are grateful to Bing Zhang, Daniel Kasen, Zong-Kai Peng, Shun-Ke Ai, Cheng-Yuan Wu, Jin-Ping Zhu, Shang-Ming Chen for insightful discussions that helped improve this work.
This work is supported by the National SKA Program of China (2020SKA0120300), the National Key R\&D Program of China (2021YFA0718500), the National Natural Science Foundation of China (grant Nos 12303047  and 12393811), Natural Science Foundation of Hubei Province (2023AFB321) and the China Manned Spaced Project (CMS-CSST-2021-A12).
\end{acknowledgements}

\appendix
\section{Equation of State of supernova ejecta} \label{sec:EOS}

The equation of state for freshly shocked, optically thick supernova ejecta is a key component in modeling supernova light curves. It is usually written as the sum of ideal‐gas and radiation contributions:
\begin{equation}
  P = P_{\mathrm{gas}} + P_{\mathrm{rad}} = \frac{\rho k_B T}{\mu m_u} + \frac{a T^4}{3},
\end{equation}
where $\mu$ is the mean molecular weight and $m_u$ is the atomic mass unit, $k_{B}$ is the Boltzmann constant, and  $a$ is the radiation constant.

The average density is approximately:
\begin{equation}
  \rho \approx \frac{3 M_{\mathrm{ej}}}{4 \pi R^3} \approx 5 \times 10^{-13}  \left( \frac{M_{\mathrm{ej}}}{M_{\odot}} \right) \left( \frac{R}{10^{15} \, \text{cm}} \right)^{-3} \, \text{g cm}^{-3},
\end{equation}
where \( M_{\mathrm{ej}} \) is the ejecta mass and \( R \) is the radius.

The strong shock approximately divides the total energy of the supernova ejecta into thermal and kinetic components  ($E_{\rm{Th}}$ and $E_{\rm K}$) forms, which is estimated by:
\begin{equation}
  E_{\mathrm{SN}} \approx 2 a T^4 \left( \frac{4 \pi R^3}{3} \right),
\end{equation}
which leads to the temperature:
\begin{equation}
  T \simeq 6.3 \times 10^4  \left( \frac{E_{\mathrm{SN}}}{10^{51} \, \text{erg}} \right)^{-\frac{1}{4}} \left( \frac{R}{10^{15} \, \text{cm}} \right)^{-\frac{3}{4}} \, \text{K}.
\end{equation}

For typical supernova values, the ratio of radiation pressure to gas pressure is:
\begin{equation}
  \frac{P_{\mathrm{rad}}}{P_{\mathrm{gas}}} \simeq 1.6 \times 10^4 \left( \frac{E_{\mathrm{SN}}}{10^{51} \, \text{erg}} \right)^{\frac{1}{4}} \left( \frac{R}{10^{15} \, \text{cm}} \right)^{\frac{1}{4}} \left( \frac{M_{\mathrm{ej}}}{M_{\odot}} \right)^{-1}.
\end{equation}
This ratio is significantly greater than 1 for typical values, indicating that radiation pressure strongly dominates over gas pressure in optically thick supernova ejecta. This dominance is a crucial simplification often employed in the mathematical frameworks used for modeling supernova light curves.

\section{Numerical Implementation of Time-Dependence Radiation Diffusion }
\label{sec:Numerical Implementation}

To generate theoretical light curves, it is crucial to solve the partial
differential equation governing the dimensionless internal energy density $e
(x, y)$. Accurately solving this equation enables us to predict the evolution
of energy, and consequently the observed luminosity. For computational
efficiency, we simplify the mathematical formulation of the governing
equations, boundary conditions, and initial conditions, which can be
summarized as follows:
\begin{equation}
\left\{
\begin{alignedat}{2}
\frac{\partial e(x, y)}{\partial y} 
&= \frac{1}{x^2} \frac{\partial}{\partial x} \left[ D(x, y) \frac{\partial e(x, y)}{\partial x} \right] + S(x, y), 
\quad && \text{for } x_{\min} < x < x_{\max},\ y > y_{\min}, \\
\frac{\partial e(x, y)}{\partial x} 
&= f_{\mathrm{ib}}(y), 
\quad && \text{for } x = x_{\min},\ y > y_{\min}, \\
e(x, y) 
&= f_{\mathrm{ob}}(y) \frac{\partial e(x, y)}{\partial x}, 
\quad && \text{for } x = x_{\max},\ y > y_{\min}, \\
e(x, y) 
&= f_{\mathrm{initial}}(x), 
\quad && \text{for } x_{\min} \leq x \leq x_{\max}, y=y_{\min}.
\end{alignedat}
\right.
\end{equation}

The Crank-Nicolson scheme is a finite difference method commonly used for numerically solving partial differential equations, especially in heat conduction problems. It is an implicit scheme, known for its unconditional stability, and employs time-centered averaging to enhance accuracy.

We discretize the temporal variable $y$ into a uniform grid with step size
$\Delta y$. The discrete time points, denoted by index $n$, are defined as:
\begin{equation}
y_n = y_{\min} + n \Delta y, ~ n = 0, 1, 2, \ldots, N_y,
\end{equation}
where $y_n$ represents the time grid points.

Similarly, the spatial domain between $x_{\min}$ and $x_{\max}$ is discretized
using a uniform spatial grid with step size $\Delta x$. The discrete spatial
points, denoted by index $j$, are defined as:
\begin{equation}
x_j = x_{\min} + j \Delta x, ~ j = 0, 1, 2, \ldots, N_x,
\end{equation}
where $x_j$ represents the spatial grid points.

We denote the numerical approximation to $e (x_j, y_n)$ by $e_j^n$. The Crank--Nicolson scheme applies centered differences to the spatial derivatives and employs a time-centered approach by averaging over two time levels.
Substituting the discretized time and space derivatives into the governing equation yields the following discrete system:
\begin{equation}
\begin{aligned}
    \frac{e_{j}^{n+1}-e_{j}^{n}}{\Delta y} &= \frac{1}{x_j^2 2\Delta x^2} \left[ D_{j+\frac{1}{2}}^{n+1}(e_{j+1}^{n+1}-e_j^{n+1}) - D_{j-\frac{1}{2}}^{n+1}(e_j^{n+1}-e_{j-1}^{n+1}) \right] \\
    &\quad + \frac{1}{x_j^2 2\Delta x^2} \left[ D_{j+\frac{1}{2}}^{n}(e_{j+1}^n-e_j^n) - D_{j-\frac{1}{2}}^{n}(e_j^n-e_{j-1}^n) \right] \\
    &\quad + \frac{1}{2}(S_j^{n+1} + S_j^n),
\end{aligned}
\end{equation}
where the averaged diffusion coefficients at half-grid points are defined by
\begin{equation}
    D_{j + \frac{1}{2}}\equiv  \frac{1}{2} (D_{j + 1} + D_j) ~ \text{ and } ~ D_{j - \frac{1}{2}}\equiv   \frac{1}{2} (D_j + D_{j - 1}),
\end{equation}
with $j = 1, 2, \ldots, N_x-1$.

We assume that the solution at the time level $n$ ($e_j^n$) is known,
while all quantities at the new time level $n + 1$ are unknown. The objective
is thus to advance the solution from $e_j^n$ at $y_n$ to $e_j^{n + 1}$ at
$y_{n + 1} = y_n + \Delta y$.

After rearranging terms, multiplying by $\Delta y$, and collecting unknown
terms (those at the time level $n + 1$) on the left-hand side, we obtain the
linear algebraic system:
\begin{equation} \label{Eq:Diff}
-\mu_j(D_j^{n+1}+D_{j-1}^{n+1})e_{j-1}^{n+1}+[1+\mu_j(D_{j+1}^{n+1}+2D_j^{n+1}+D_{j-1}^{n+1})]e_j^{n+1}-\mu_j(D_{j+1}^{n+1}+D_j^{n+1})e_{j+1}^{n+1}=b_j,
\end{equation}
where we have introduced the definitions:
\begin{equation}
    \mu_j \equiv \frac{1}{4} \frac{\Delta y}{x_j^2 (\Delta x)^2},
\end{equation}
and the known terms are grouped into $b_j$, defined by
\begin{equation}
    b_j \equiv \mu_j (D_j^n + D_{j - 1}^n) e_{j - 1}^n + [1 - \mu_j (D_{j +
   1}^n + 2 D_j^n + D^n_{j - 1})] e^n_j + \mu_j (D_{j + 1}^n + D_j^n) e_{j +
   1}^n + \frac{1}{2} (S_j^{n + 1} + S_j^n) \Delta y. 
\end{equation}

The unknown variables $e_{j - 1}^{n + 1}$, $e_j^{n + 1}$, and $e_{j + 1}^{n +
1}$ at the new time level $n + 1$ are coupled through the linear algebraic
system shown in Eq.(\ref{Eq:Diff}). This system can be simplified into the following tridiagonal form:
\begin{equation}
    A_{j, j - 1} e_{j - 1}^{n + 1} + A_{j, j} e_j^{n + 1} + A_{j, j + 1} e^{n +
1}_{j + 1} = b_j,
\end{equation}
where the coefficients are defined by
\begin{equation}
\begin{aligned}A_{j,j-1}&\equiv-\mu_j(D_j^{n+1}+D_{j-1}^{n+1}),\\A_{j,j}&\equiv1+\mu_j(D_{j+1}^{n+1}+2D_j^{n+1}+D_{j-1}^{n+1}),\\A_{j,j+1}&\equiv-\mu_j(D_{j+1}^{n+1}+D_j^{n+1}),\end{aligned}
\end{equation}
in the equations for internal points, $j = 1, 2, \ldots, N_x-1$. 
At the inner boundary ($j = 0$), the boundary condition yields:
\begin{equation}
A_{0,0} = -1,\quad A_{0,1} = 1,\quad b_0 = \Delta x\,f_{\mathrm{ib}}^n.
\end{equation}
At the outer boundary ($j = N_x$), the boundary condition is given by:
\begin{equation}
A_{N_x, N_x - 1} = -f_{\mathrm{ob}}^n,\quad A_{N_x, N_x} = \Delta x + f_{\mathrm{ob}}^n,\quad b_{N_x} = 0.
\end{equation}
The coupled system of algebraic equations can be written on matrix form as
\begin{equation}
    A\mathbf{U}= \mathbf{b},
\end{equation}
where  $\mathbf{U} = (e_0^{n + 1}, e_1^{n + 1}, \cdots, e_{N_x}^{n + 1})$, and $\mathbf{b} = (b_0, b_1, \cdots b_{N_x})$, the matrix $A$ has the following structure:

\begin{equation}
A=\begin{pmatrix}A_{0,0}&A_{0,1}&0&\cdots&\cdots&\cdots&\cdots&\cdots&0\\A_{1,0}&A_{1,1}&A_{1,2}&\ddots&&&&&\vdots\\0&A_{2,1}&A_{2,2}&A_{2,3}&\ddots&&&&\vdots\\\vdots&\ddots&&\ddots&\ddots&0&&&\vdots\\\vdots&&\ddots&\ddots&\ddots&\ddots&\ddots&&\vdots\\\vdots&&&0&A_{j,j-1}&A_{j,j}&A_{j,j+1}&\ddots&\vdots\\\vdots&&&&\ddots&\ddots&\ddots&\ddots&0\\\vdots&&&&&\ddots&\ddots&\ddots&A_{N_{x}-1,N_{x}}\\0&\cdots&\cdots&\cdots&\cdots&\cdots&0&A_{N_{x},N_{x}-1}&A_{N_{x},N_{x}}\end{pmatrix}.
\end{equation}
At each time step both the right–hand side vector $\mathbf{b}$ and the tridiagonal coefficient matrix $A$ of size $(N_x+1)\times(N_x+1)$
must be refreshed.  Such a tridiagonal system can be efficiently solved using standard numerical methods, for example, the Thomas algorithm. Alternatively, numerical libraries such as the \texttt{scipy.sparse.linalg.spsolve} function from \texttt{SciPy} can exploit the sparse structure of the matrix $A$, providing an efficient Gaussian elimination-based solution.

Given a set of ejecta parameters and heating‑source parameters, the
dimensionless internal‑energy density \(e_j^n\) is evaluated numerically at
each spatial grid point \(x_j\) and time level \(y_n\).
The bolometric luminosity at the corresponding observer time is obtained from the diffusive flux exiting the outermost grid cell:
\begin{equation}
L_{\mathrm{bol}}(t)
   \;=\;
   -\,L_{0}\,
   \frac{x_{\max}^{2}}{\eta_{\mathrm{ej}}\!\bigl(x_{\max}\bigr)}
   \,
   \frac{e^{\,n}_{N_x} - e^{\,n}_{N_x-1}}{\Delta x}.
\end{equation}

\section{Integrating factor method for solving linear first-order ordinary differential equations}
\label{Sec:Integrating_factor}

To solve the first-order linear ordinary differential equation given by:
\begin{equation}
 \frac{\mathrm{d} \phi}{\mathrm{d} y} + \alpha \left( 1 + \frac{y}{y_{\mathrm{ex}}} \right) \phi - b\left( 1 + \frac{y}{y_{\mathrm{ex}}} \right) f_{\mathrm{heat}}(y) = 0
\end{equation}
we first rewrite the equation in the standard linear ordinary differential equation form:
\begin{equation}
\frac{\mathrm{d} \phi}{\mathrm{d} y} + P (y) \phi = Q (y),
\end{equation}
where
\begin{equation} \label{Eq:Py}
P (y) = \alpha \left( 1 + \frac{y}{y_{\mathrm{ex}}} \right),
\end{equation}
and
\begin{equation}
Q (y) = b\left( 1 + \frac{y}{y_{\mathrm{ex}}} \right) f_{\mathrm{heat}}(y).
\end{equation}
To solve this, the integrating factor method is employed. The integrating factor $I(y)$ is defined as:
\begin{equation}
I (y) \equiv  e^{\int P (y) \mathrm{d} y}.
\end{equation}
Substituting $P(y)$ from Equation~(\ref{Eq:Py}), the integrating factor becomes:
\begin{equation}
I (y) = e^{\alpha \left( y + \frac{y^2}{2 y_{\mathrm{ex}}} \right)}.
\end{equation}
Multiplying the standard form of the differential equation by the integrating factor yields:
\begin{equation}
e^{\alpha \left( y + \frac{y^2}{2 y_{\mathrm{ex}}} \right)} \frac{\mathrm{d} \phi}{\mathrm{d} y} + e^{\alpha \left( y + \frac{y^2}{2 y_{\mathrm{ex}}} \right)} \alpha \left( 1 + \frac{y}{y_{\mathrm{ex}}} \right) \phi = e^{\alpha \left( y + \frac{y^2}{2 y_{\mathrm{ex}}} \right)} b\left( 1 + \frac{y}{y_{\mathrm{ex}}} \right) f_{\mathrm{heat}}(y) .
\end{equation}
The left-hand side of this equation is the derivative of the product of the integrating factor and $\phi (y)$:
\begin{equation}
\frac{\mathrm{d}}{\mathrm{d} y} \left[ e^{\alpha \left( y + \frac{y^2}{2 y_{\mathrm{ex}}} \right)} \phi \right] = be^{\alpha \left( y + \frac{y^2}{2 y_{\mathrm{ex}}} \right)} \left( 1 + \frac{y}{y_{\mathrm{ex}}} \right) f_{\mathrm{heat}}(y) .
\end{equation}
Integrating both sides with respect to $y$ gives:
\begin{equation}
e^{\alpha \left( y + \frac{y^2}{2 y_{\mathrm{ex}}} \right)} \phi = \int_{0}^{y} be^{\alpha \left( y' + \frac{y'^2}{2 y_{\mathrm{ex}}} \right)}  \left( 1 + \frac{y'}{y_{\mathrm{ex}}} \right) f_{\mathrm{heat}} (y') \mathrm{d} y' + C,
\end{equation}
where $C$ is the constant of integration, which would be determined by an initial condition for $\phi (y)$.

Finally, solving for $\phi (y)$, we obtain:
\begin{equation}
\phi (y) = b e^{- \alpha \left( y + \frac{y^2}{2 y_{\mathrm{ex}}} \right)} \int_{0}^{y} e^{\alpha \left( y' + \frac{y'^2}{2 y_{\mathrm{ex}}} \right)}\left( 1 + \frac{y'}{y_{\mathrm{ex}}} \right) f_{\mathrm{heat}} (y') \mathrm{d} y' + C e^{- \alpha \left( y + \frac{y^2}{2 y_{\mathrm{ex}}} \right)}.
\end{equation}

Similarly, for the equation,
\begin{equation}
\frac{\mathrm{d} L}{\mathrm{d} t} = \frac{t}{\tau_{\mathrm{d}}^2} (L_{\mathrm{heat}} - L),
\end{equation}
using the same method with the initial condition $L(0) = 0$, the solution is obtained as
\begin{equation}
L (t) = e^{- \frac{t^2}{2 \tau_{\mathrm{d}}^2}} \int_0^t \frac{t'}{\tau_{\mathrm{d}}^2} L_{\mathrm{heat}} (t') e^{\frac{{t'}^2}{2 \tau_{\mathrm{d}}^2}} \mathrm{d} t'.
\end{equation}

\section{NOTATION LIST}
\label{sec:NotationList}
The notation we used is listed in Table.(\ref{tab:notation}).
\startlongtable
\begin{deluxetable*}{lll}
\tablecaption{Physical symbols used in this work\label{tab:notation}}
\tablehead{\colhead{Symbol} & \colhead{Definition / description} & \colhead{Unit}}
\tabletypesize{\small}
\tablewidth{0pt}
\startdata
\(t\)                        & Physical time                                       & s \\
\(r\)                        & Radial coordinate (Eulerian)                        & cm \\
\(m\)                        & Lagrangian mass coordinate                          & g \\
\(R_{0}\)                    & Progenitor radius at shock breakout                 & cm \\
\(v_{\max}\)                 & Maximum ejecta velocity                             & cm\,s\(^{-1}\) \\
\(R_{\max}=R_{0}+v_{\max}t\) & Outer radius (homologous expansion)                 & cm \\
\(x = r/R_{\max}\)           & Dimensionless radius                                & dimensionless \\
\(y = t/t_{\mathrm{diff}}\)  & Dimensionless time                                  & dimensionless \\[2pt]
\(t_{\mathrm{diff}} =3\kappa\rho_{0}R_{0}^{2}/c\)        & Diffusion timescale & s \\[2pt]
\(\rho(r,t)\)               & Density                                             & g\,cm\(^{-3}\) \\
\(\rho_{0}\)                & Characteristic density                              & g\,cm\(^{-3}\) \\
\(\eta_{\mathrm{ej}}(x)\)    & Normalised density profile                          & dimensionless \\
\(P\)                        & Pressure                                            & dyn\,cm\(^{-2}\) \\
\(E\)                        & Specific internal energy                            & erg\,g\(^{-1}\) \\
\(u=\rho E\)                & Internal-energy density                             & erg\,cm\(^{-3}\) \\
\(u_{0}\)                    & Characteristic energy density                       & erg\,cm\(^{-3}\) \\[2pt]
\(F\)                        & Radiative flux                                      & erg\,cm\(^{-2}\)s\(^{-1}\) \\
\(L\)                        & Local (shell) luminosity                            & erg\,s\(^{-1}\) \\
\(L_{\mathrm{bol}}\)         & Bolometric luminosity at the surface                & erg\,s\(^{-1}\) \\
\(L_{0}=4\pi R_{0}^{3}u_{0}/t_{\mathrm{diff}}\) & Characteristic luminosity        & erg\,s\(^{-1}\) \\
\(c\)                        & Speed of light                                      & cm\,s\(^{-1}\) \\[2pt]
\(\kappa\)                   & Gray opacity                                        & cm\(^{2}\)g\(^{-1}\) \\
\(\varepsilon_{\mathrm{heat}}(r,t)\) & Heating rate per unit mass                  & erg\,g\(^{-1}\)s\(^{-1}\) \\
\(\varepsilon_{\mathrm{heat}0}\)    & Heating-rate amplitude                       & erg\,g\(^{-1}\)s\(^{-1}\) \\
\(\xi_{\mathrm{heat}}(r)\)   & Radial heating profile                              & dimensionless \\
\(f_{\mathrm{heat}}(t)\)     & Temporal heating factor                             & dimensionless \\
\(x_{\mathrm{heat}}\)        & Mixing radius of \(^{56}\mathrm{Ni}\) (dimensionless) & dimensionless \\
\(D(x,y)\)                   & Dimensionless diffusion coefficient                 & dimensionless \\
\(S(x,y)\)                   & Dimensionless source term                           & dimensionless \\[2pt]
\(M_{\mathrm{ej}}\)          & Ejected mass                                        & g (or \(M_\odot\)) \\
\(M_{\mathrm{Ni}}\)          & Synthesised \(^{56}\mathrm{Ni}\) mass               & g (or \(M_\odot\)) \\
\(E_{\mathrm{K}}\)           & Kinetic energy                                      & erg \\
\(E_{\mathrm{Th}}\)          & Thermal energy                                      & erg \\
\(E_{\mathrm{Th,in}}\)       & Initial thermal energy                              & erg \\
\(E_{\mathrm{SN}}\)          & Total explosion energy                              & erg \\
\(E_{\mathrm{int}}\)         & One-zone internal energy                            & erg \\[2pt]
\(E_{\mathrm{rot}}\)         & Magnetar rotational energy                          & erg \\
\(I_{\mathrm{NS}}\)          & Neutron-star moment of inertia                      & g\,cm\(^{2}\) \\
\(P_{\mathrm{i}}\)           & Magnetar initial spin period                        & s \\
\(B_{\mathrm{d}}\)           & Dipole magnetic-field strength                      & G \\
\(t_{\mathrm{sd}}\)          & Spin-down timescale                                 & s \\
\(L_{\mathrm{sd}}(t)\)       & Spin-down luminosity                                & erg\,s\(^{-1}\) \\
\(L_{\mathrm{engine}}(t)\)   & Generic central-engine luminosity                   & erg\,s\(^{-1}\) \\[2pt]
\(v_{\mathrm{ej}}=\sqrt{2E_{\mathrm{K}}/M_{\mathrm{ej}}}\) & Characteristic ejecta velocity & cm\,s\(^{-1}\) \\[2pt]
\(\tau_{0}\)                 & Initial optical depth                               & dimensionless \\
\(\tau(t)\)                  & Optical depth at time \(t\)                         & dimensionless \\
\(R_{\mathrm{ph}}\)          & Photospheric radius (\(\tau=2/3\))                  & cm \\
\(x_{\mathrm{ph}}=R_{\mathrm{ph}}/R_{\max}\) & Dimensionless photospheric radius & dimensionless \\[2pt]
\(t_{\mathrm{ex}}=R_{0}/v_{\max}\) & Expansion timescale                       & s \\
\(\tau_{\mathrm{m}}\)         & Arnett-model diffusion time                         & s \\
\(\tau_{\mathrm{d}}\)         & One-zone diffusion time                             & s \\[2pt]
\(a\)                        & Radiation constant        & erg\,cm\(^{-3}\)K\(^{-4}\) \\
\(k_{\mathrm{B}}\)           & Boltzmann constant                                  & erg\,K\(^{-1}\) \\
\(\mu\)                      & Mean molecular weight                               & dimensionless \\
\(m_{\mathrm{u}}\)         & Atomic mass unit                                    & g \\
\(T_{\mathrm{Ni}}\)          & \(^{56}\mathrm{Ni}\) synthesis temperature           & K \\
\(R_{\mathrm{Ni}}\)          & Radius where \(T\ge T_{\mathrm{Ni}}\)               & cm \\[2pt]
\(f_{\mathrm{ib}}(y)\)       & Inner-boundary flux factor                          & dimensionless \\
\(f_{\mathrm{ob}}(y)\)       & Outer-boundary flux factor                          & dimensionless \\[2pt]
\(I_{\mathrm{M}}\)           & Mass integral \(=\int_0^1 x^{2}\eta_{\mathrm{ej}}\,dx\)         & dimensionless \\
\(I_{\mathrm{K}}\)           & Kinetic integral \(=\int_0^1 x^{4}\eta_{\mathrm{ej}}\,dx\)      & dimensionless \\
\(I_{\mathrm{Th}}\)          & Thermal integral \(=\int_0^1 x^{2}e(x,0)\,dx\)                  & dimensionless \\
\(I_{\tau}\)                 & Optical-depth integral \(=\int_0^1 \eta_{\mathrm{ej}}\,dx\)      & dimensionless \\[2pt]
\(L_{\mathrm{peak}}\)        & Peak bolometric luminosity                          & erg\,s\(^{-1}\) \\
\enddata
\end{deluxetable*}

\bibliography{sample631}{}
\bibliographystyle{aasjournal}

\end{document}